\documentclass[%
 reprint,
superscriptaddress,
 amsmath,amssymb,
 aps,
]{revtex4-2}

\usepackage{graphicx}
\usepackage{dcolumn}
\usepackage{bm}
\usepackage{multirow}

\begin{document}

\preprint{APS/123-QED}

\title{The extraction of higher-order radial moments of nuclear charge density from muonic atom spectroscopy }

\author{Hui Hui Xie}
 \affiliation{College of Physics, Jilin University, Changchun 130012, China}
\author{Jian Li}\email{jianli@jlu.edu.cn}
\affiliation{College of Physics, Jilin University, Changchun 130012, China}

\author{Haozhao Liang}\email{haozhao.liang@phys.s.u-tokyo.ac.jp}
\affiliation{Department of Physics, Graduate School of Science,	The University of Tokyo, Tokyo 113-0033, Japan}
\affiliation{RIKEN Interdisciplinary Theoretical and Mathematical Sciences Program, Wako 351-0198, Japan}

\date{\today}

\begin{abstract}
Muonic atom transitions have been measured for almost all stable nuclei to extract nuclear structure properties, including nuclear charge radii and quadrupole moment. 
To investigate the possibilities of extracting higher-order radial moments of nuclear charge density 
from muonic atom spectroscopy, a theory-to-theory benchmark analysis based on a model-independent density distribution, i.e., the Fourier-Bessel series expansion instead of two-parameter Fermi distribution, is performed by taking $^{208}$Pb as an example, where nuclear charge density obtained from the relativistic continuum Hartree-Bogoliubov calculation is used as the benchmark. 
It is found that the extractions of the higher-order moments, i.e., the fourth and sixth moments in addition to the second moment are feasible with high accuracy.
Moreover, the charge form factor in the low-$q$ region can also be well extracted.

\end{abstract}

\maketitle


\section{Introduction}

Nuclear charge radii $\sqrt{\langle r^2_c\rangle}$ have been well measured in recent decades by several electromagnetic methods as shown in the compilations~\cite{ANGELI2004185,ANGELI201369,LI2021101440}.
For example, the charge radii of stable nuclei can be extracted from the parameterized charge density in terms of Fourier-Bessel (FB) series expansion, with coefficients determined from the cross-section data in the elastic electron scattering experiments~\cite{RevModPhys.28.214,DREHER1974219,DEVRIES1987495}.
The charge radii of stable nuclei have also been extracted by the muonic transitions revealed from the muonic X-rays, based on the two-parameter Fermi (2pf) distribution~\cite{ENGFER1974509,PhysRevC.23.533,PhysRevC.37.2821,SCHALLER1978225,PhysRevC.18.1474,PhysRevC.39.195,Knecht2020}.
In contrast, the charge radii of unstable isotopes can only be measured from the laser spectroscopy, which provides information on the changes in mean squared radii $\delta\langle r^2_c\rangle$~\cite{CAMPBELL2016127,PhysRevC.94.054321,Koszorus2021}. 

In laser spectroscopy, the changes in mean-square charge radii are deduced from the isotope shifts, which are extracted from the hyperfine spectra on the atomic and ionic transitions~\cite{PhysRevC.94.054321,S_A_Blundell_1987}.
As one of the components of isotope shifts, the field shift depends on the changes in nuclear charge distribution and hence is responsible for the extraction of $\delta \langle r^2_c\rangle$~\cite{S_A_Blundell_1987}.
In addition to $\delta\langle r^2_c\rangle$, it is demonstrated by Papoulia \textit{et al}.~\cite{PhysRevA.94.042502} that the extraction of changes in the fourth moment $\delta \langle r^4_c\rangle$ for heavy nuclei is possible through an improved description of the field shift, since such information is inherently contained in the field shift. From this viewpoint, it should also be possible to extract the high-order moments of nuclear charge density, such as the fourth and sixth moments, from muonic atom spectroscopy, considering that the mass of a muon is about $207$ times heavier than that of an electron.

As shown in one of our previous studies~\cite{xie2023revisiting},
the influence of model dependency induced by employing the 2pf distribution on the extraction of charge radii is quite small.
However, the extraction of additional nuclear information, such as higher-order moments of charge density, could be limited by such a model dependency.
Therefore, it is worthwhile to investigate what kind of information can be extracted from the muonic atom spectroscopy with the parameterization of charge density in terms of a model-independent distribution, such as the Fourier-Bessel series expansion~\cite{DREHER1974219}.
For this reason, a similar theory-to-theory benchmarking analysis will be made in the present study.
It is also demonstrated in Ref.~\cite{xie2023revisiting} that the charge radii of heavy nuclei can be extracted more accurately than those of the light nuclei.
Thus, the double-magic nucleus $^{208}$Pb will be taken as an example here, while the theoretical charge density obtained from the covariant density functional theory (CDFT) will be used as the benchmark.

In recent decades, the CDFT has attracted extensive attention on account of its successful descriptions of many nuclear phenomena~\cite{RING1996193,MENG2006470,Meng_2015,VRETENAR2005101,NIKSIC2011519,meng2013progress,SHARMA19939,PhysRevC.107.054307,LIANG20151,PhysRevLett.91.262501,liang2010spin,Li_2020,Li2011,10.1143/PTP.125.1185,10.1143/PTPS.196.400,PhysRevC.88.064307,PhysRevLett.71.3079,PhysRevLett.107.122501}. 
To provide a unified and self-consistent treatment of the continuum, the mean-field potentials, and the pairing correlations, the relativistic continuum Hartree-Bogoliubov (RCHB) theory~\cite{PhysRevLett.77.3963,Meng1998NPA} has been developed by extending the CDFT with the Bogoliubov transformation in the coordinate representation, and it has achieved great success in various aspects~\cite{PhysRevLett.77.3963,PhysRevLett.80.460,PhysRevC.65.041302,Shuang_Quan_2002,MENG19981,MENG2002209}.
In this work, the theoretical charge density of $^{208}$Pb is provided from the RCHB calculations.

The paper is organized as follows.
In Sec.~\ref{sec2}, we introduce the RCHB method to construct the nuclear charge density and the Dirac equation for muonic atom with the FB charge distribution.
In Sec.~\ref{sec3}, we illustrate an iteration scheme constructed to yield the best-fit FB distribution and discuss the possibilities of extracting nuclear information, including the second, fourth, and sixth moments, as well as the detailed charge density and the corresponding charge form factor (FF).
Finally, the summary is presented in Sec.~\ref{sec4}.

\section{Theoretical framework}\label{sec2}

\subsection{Nuclear charge density from RCHB theory}

In this study, the RCHB theory constructed with the contact interaction in the point-coupling representation between nucleons is adopted. 
The details of the RCHB theory 
can be found in Refs.~\cite{PhysRevC.82.054319,Xia2018}. In the following, we briefly introduce the theoretical framework 
of RCHB theory.

Starting from the Lagrangian density, the energy density functional of the nuclear system can be constructed under the mean-field and no-sea approximations.
By minimizing the energy density functional with respect to the
densities, one obtains the Dirac equation for nucleons within the
relativistic mean-field framework~\cite{Meng2015}. The relativistic Hartree-Bogoliubov model provides a unified description of both the mean field and the pairing correlation.

In the RCHB theory, the proton and neutron densities can be constructed by quasiparticle wave functions,
\begin{equation}\label{eq:rhov}
	\rho_{\tau}(\bm r)   =\sum_{k\in\tau } V_k^{\dagger}(\bm r)V_k(\bm r)
\end{equation}
with $\tau\in\{p,n\}$.
The nuclear charge density then includes the contributions from the point neutron density, the proton and neutron spin-orbit densities, and the single-proton and single-neutron charge densities, in addition to that from the point proton
density~\cite{Friar1975,PhysRevC.62.054303,PhysRevC.103.054310,kurasawa2019n}. The relativistic nuclear charge density is finally written as
\begin{equation}\label{eq-rhoc}
	\rho_c(r)=\sum_{\tau}\left[\rho_{c\tau}(r)+W_{c\tau}(r)\right],
\end{equation}
where
\begin{align}
	&\rho_{c\tau}(r)=\frac1r\int_0^\infty x\rho_\tau(x)\left[g_\tau(|r-x|)-g_\tau(r+x)\right]dx,\\
	&W_{c\tau}(r)=\frac1r\int_0^\infty xW_\tau(x)\left[f_{2\tau}(|r-x|)-f_{2\tau}(r+x)\right]dx.
\end{align}
Here, $\rho_\tau(r)$ is the point nucleon density in Eq.~(\ref{eq:rhov}) and $W_\tau(r)$ is the spin-orbit density given in Refs.~\cite{PhysRevC.62.054303,kurasawa2019n}. The functions $g_\tau(x)$ and $f_{2\tau}(x)$ are given by
\begin{eqnarray}
	g_\tau(x)=\frac1{2\pi}\int_{-\infty}^{\infty}e^{iqx}\, G_{E\tau}(\bm q^2)dq,\\
	f_{2\tau}(x)=\frac{1}{2\pi}\int_{-\infty}^{\infty}e^{iqx}\, F_{2\tau}(\bm q^2)dq,
\end{eqnarray}
in which $G_{E\tau}$ and $F_{2\tau}$ denote the electric Sachs and Pauli form factors of a nucleon, respectively. The following forms~\cite{PhysRevC.62.054303} are used in this study,
\begin{align}\label{eq-form-factor}
	G_{Ep}(q^2) &=\frac{1}{\left(1+r_p^2\bm q^2/12\right)^2}, \nonumber\\
	G_{En}(q^2) &=\frac{1}{\left(1+r_+^2\bm q^2/12\right)^2}-\frac{1}{\left(1+r_-^2\bm q^2/12\right)^2},\nonumber\\
	F_{2p}(q^2) &=\frac{G_{Ep}}{1+\bm q^2/4M_p^2},\nonumber\\
	F_{2n}(q^2) &=\frac{G_{Ep}-G_{En}/\mu_n}{1+\bm q^2/4M_n^2},
\end{align}
with the proton charge radius $r_p= 0.8414$~fm~\cite{RevModPhys.93.025010} and $r^2_\pm= r_{\mathrm{av}}^2\pm \frac12 \langle r_n^2\rangle $, where $r_{\mathrm{av}}^2= 0.81$~fm$^2$ is the average of the mean-square radii for positive and negative charge distributions, while $\langle r_n^2\rangle =-0.11$~fm$^2$~\cite{atac2021measurement} is the mean squared charge radius of a neutron. See Ref.~\cite{PhysRevA.107.042807,xie2023impact} for more details.

\subsection{Dirac equation for muonic atom}

The Dirac equation for the muonic atom reads
\begin{equation}\label{eq-dirac-one-electron}
	\left[\bm\alpha\cdot \bm p+\beta M_r+V(\bm r)\right]\psi_k(\bm r)=\varepsilon_k \psi_k(\bm r),
\end{equation}
where the eigenenergy $\varepsilon_k$ of state $k$ includes both the muonic atom energy level $E_k$ and the reduced mass $M_r$, i.e., 
\begin{equation}\label{reduced-mass}
	\varepsilon_k \simeq E_k+M_r \quad \mbox{with} \quad  M_r=\frac{M_Am_\mu}{M_A+m_\mu}.
\end{equation}
Note that Eq.~\eqref{reduced-mass} holds exactly only at the non-relativistic limit, while it is a good enough approximation in the present calculations.
Following the treatment of our previous work~\cite{xie2023revisiting}, the relativistic recoil correction~\cite{PhysRevA.96.032510} is not considered here for simplifying the calculation.
The mass of a muon $m_\mu=206.7682830\,m_e$ is adopted~\cite{RevModPhys.93.025010} and the nuclear mass $M_A$ is taken from Ref.~\cite{Wang_2021}. Since the electrostatic potential $V(r)$ made by the atomic nucleus is spherically symmetric, the eigenvalue function can be written as
\begin{equation}\label{eq-dirac-s}
	\psi_{n\kappa m}(\bm r)=\frac1r\begin{pmatrix}
		iP_{n\kappa}(r)Y_{jm}^l(\theta,\varphi)\\
		Q_{n\kappa}(r)\left(\bm\sigma\cdot\bm{\hat r}\right)Y_{jm}^l(\theta,\varphi)
	\end{pmatrix},
\end{equation}
where $P_{n\kappa}(r)$ and $Q_{n\kappa}(r)$ are its large and small components, respectively.

The radial Dirac equation for the one-muon system can then be written in the form of
\begin{align}\label{eq-dirac}
	\begin{pmatrix}
		V(r) & -\left(\frac{ d}{ dr}-\frac{\kappa}{ r}\right)\\
		\left(\frac{ d}{ dr}+\frac \kappa r\right) &  V(r)-2M_r
	\end{pmatrix}
	\begin{pmatrix}
		P_{n\kappa}(r)\\Q_{n\kappa}(r)
	\end{pmatrix}&=
	E_{n\kappa}\begin{pmatrix}
		P_{n\kappa}(r)\\Q_{n\kappa}(r)
	\end{pmatrix},
\end{align}
where the mass term $M_r$ is subtracted on both sides of the equation.  

The electrostatic potential $V(r)$ is obtained via~\cite{engel2002relativistic}
\begin{equation}\label{eq-v}	V(r)=-4\pi\alpha\left[\int_0^r\rho_c(r')\frac{r'^2}{r}dr'+\int_r^\infty\rho_c(r')r'dr'\right],
\end{equation}
where $\alpha$ is the fine structure constant.
In the present work, the muonic energy levels are evaluated numerically.

\subsection{Model-independent Fourier-Bessel analysis}

The Fourier-Bessel series expansion was introduced by Dreher \textit{et al}.~\cite{DREHER1974219} to perform a model-independent analysis of measured electron scattering data, i.e., the differential cross section $d\sigma(E,\theta)/d\Omega$ for the elastic and inelastic scatterings of an electron of energy $E$ through an angle $\theta$. In the first Born approximation, the elastic cross section can be written as 
\begin{equation}
	\frac{d\sigma}{d\Omega}=\left(\frac{Ze^2}{2E}\right)\frac{\cos^2\frac{\theta}{2}}{\sin^4\frac{\theta}{2}}\left|F_c(q)\right|^2,
\end{equation}
where $q=(2E/\hbar c) \sin(\theta/2)$ is the momentum transfer.
The charge form factor $F_c(q)$ can be regarded as the representation of charge density in momentum space, and it is given by a Fourier-Bessel transformation of charge density with the spherical symmetry imposed,
\begin{equation}\label{eq-Fc-rhoc}
	F_c(q)=\frac{4\pi}{Z}\int_0^\infty\rho_c(r)j_0(qr)r^2dr,
\end{equation}
where $j_0(x)=(\sin x)/x$ denotes the spherical Bessel function of order zero, and $Z$ denotes the proton number.

The charge distribution can then be expanded into a Fourier-Bessel series in the following form, assuming $\rho_c(r)$ to be zero beyond a certain cutoff radius $R_{\rm cut}$,
\begin{equation}\label{eq-rc-fb}
	\rho_c(r)=
	\begin{cases}
			\sum\limits_{\nu=1}^\infty a_\nu j_0(q_\nu r) &\mathrm{for}~r\le R_{\rm cut},\\0&\mathrm{for}~r>R_{\rm cut},
	\end{cases}
\end{equation}
with the normalization
\begin{equation}\label{eq-norm}
	4\pi\int_0^\infty\rho_c(r)r^2dr=4\pi\sum_{\nu=1}^\infty\frac{(-1)^{\nu+1}a_\nu R_{\rm cut}^3}{\left(\nu\pi\right)^2}=Z.
\end{equation}
The coefficients $a_\nu$ can be determined directly by the charge FF,
\begin{equation}\label{eq-a-Fc}	a_\nu=\frac{q_\nu^2}{2\pi R_\mathrm{cut}}F_c(q_\nu)~~\text{with}~~q_\nu=\frac{\nu\pi}{R_\mathrm{cut}}.
\end{equation}
The considered number of FB coefficients is related to the maximum value of the momentum transfer $q_\mathrm{max}$ as
\begin{equation}
	N_{\rm FB}=\frac{R_\mathrm{cut}q_\mathrm{max}}{\pi}.
\end{equation}
Combining Eq.~(\ref{eq-a-Fc}) with Eq.~(\ref{eq-Fc-rhoc}), the FB coefficients can be determined directly when the charge density is given.
In this way, a discrete charge density obtained from mean-field calculations can be easily parameterized through the FB series expansion, which is much more model-independent in comparison with the two-parameter Fermi distribution.

The $n$-th moment can be obtained with a finite number of FB coefficients 
\begin{align}
	R_n\equiv&\langle r^n_c\rangle=
	\frac{4\pi}{Z}\int_0^{R_\mathrm{cut}}\rho_c(r)r^2r^ndr\nonumber\\
	=&\frac{4\pi}{Z}\sum_{\nu=1}^{N_\mathrm{FB}} a_\nu\frac{R_\mathrm{cut}^{3+n}}{3+n}~_1F_2\left(\frac{3+n}{2};\frac32,\frac{5+n}{2};-\frac{\nu^2\pi^2}{4}\right),
\end{align}
whrer $_pF_q(a_1,\cdots,a_p;b_1,\cdots,b_q;z)$ is the generlized hypergeometric function.
In particular, the second, fourth, and sixth moments are given by
\begin{align}
	R_2=&\frac{4\pi}{Z}\sum_{\nu=1}^{N_\mathrm{FB}} a_\nu\frac{(\nu^2\pi^2-6)R_\mathrm{cut}^5}{(-1)^{1+\nu}\nu^4\pi^4},\\
	R_4=&\frac{4\pi}{Z}\sum_{\nu=1}^{N_\mathrm{FB}} a_\nu\frac{(\nu^4\pi^4-20\nu^2\pi^2-120)R_\mathrm{cut}^7}{(-1)^{1+\nu}\nu^6\pi^6},\\
	R_6=&\frac{4\pi}{Z}\sum_{\nu=1}^{N_\mathrm{FB}} a_\nu\frac{(\nu^6\pi^6-42\nu^4\pi^4+840\nu^2\pi^2-5040)R_\mathrm{cut}^9}{(-1)^{1+\nu}\nu^8\pi^8}.
\end{align}
The corresponding electrostatic potential $V(r)$ is then given by substituting Eq.~(\ref{eq-rc-fb}) into Eq.~(\ref{eq-v}),
\begin{equation}\label{eq-v-fb}
	V(r)=\begin{cases}
		\sum\limits_{\nu=1}^{N_\mathrm{FB}}a_\nu\omega_\nu(r)
		&\mathrm{for}~r\le R_\mathrm{cut},\\-\frac{Z\alpha}{r} &\mathrm{for}~r>R_\mathrm{cut},
	\end{cases}
\end{equation}
with the weighting function of FB coefficients defined as 
\begin{equation}\label{eq-omega}
	\omega_\nu(r)=-4\pi\alpha\frac{R_\mathrm{cut}^2\left[(-1)^{\nu+1}\nu\pi r+R_\mathrm{cut}\sin\left(\frac{\nu\pi r}{R_\mathrm{cut}}\right)\right]}{\nu^3\pi^3 r}. 
\end{equation}

In addition, a dimensionless deviation factor $\epsilon$ is used in this work to characterize the degree of deviation between two distributions,
\begin{equation}
	\epsilon=\frac{4\pi}{Z}\int_0^\infty\left|\rho(r)-\rho'(r)\right|r^2dr.	
\end{equation}

\section{Results and discussion}\label{sec3}

Following the treatment of our previous work~\cite{xie2023revisiting}, the numerical transition energies (with the absence of QED corrections) based on the charge densities obtained from the RCHB calculations are used as the pseudoexperimental data to constrain the FB coefficients.
The relativistic density functional PC-PK1~\cite{PhysRevC.82.054319}, which provides one of the best density-functional descriptions for nuclear properties~\cite{RN7,PhysRevC.91.027304,PhysRevC.86.064324,YAO2013459,LI2013866,LI2012470,PhysRevC.88.057301,Wang_2015}, is employed. The box size $R_{\rm box}=20$~fm, the mesh size $\Delta r=0.1$~fm, and the angular momentum cutoff $J_{\rm max}=19\hbar/2$ are used in the RCHB calculations. More numerical details can be found in Ref.~\cite{Xia2018}.

The Dirac equation for the muon is solved using the generalized pseudospectral (GPS) method, whose powerful performance has been shown in, e.g., Refs.~\cite{canuto2007spectral,PhysRevA.104.022801,https://doi.org/10.1002/qua.26653,Jiao_2021}. Using the electrostatic potential in Eq.~(\ref{eq-v-fb}), the calculations of eigenenergies for the ground state and the lower-lying excited states can be converged quickly.
The number of FB coefficients is taken as $N_{\rm FB}=15$ to achieve the model independence. 

To investigate how much information on the nuclear structure can be extracted from the muonic atom spectroscopy, we perform a theory-to-theory benchmarking analysis by taking $^{208}$Pb as an example, in which a set of best-fit FB coefficients (denoted by ``FB($\Delta E$)") that gives almost identical transition energies with the targeted values is searched.
The root-mean-square deviation (RMSD) for muonic transition energies between the target (i.e., the RCHB charge density) and the FB distribution is defined as
\begin{equation}
	\chi=\left[\frac1N\sum_i\left(\Delta E_i^{\rm FB}-\Delta E_i^{\rm RCHB}\right)^2\right]^{1/2},
\end{equation}
where $\Delta E_i^{\rm FB}$ and $\Delta E_i^{\rm RCHB}$ is the $i$-th transition energy based on the FB distribution and the RCHB charge density, respectively, and $N$ denote the total number of considered transitions. Twelve transitions are considered in the present work, i.e., $2p_{1/2}$-$1s_{1/2}$, $2p_{3/2}$-$1s_{1/2}$, $3d_{3/2}$-$1s_{1/2}$, $3d_{5/2}$-$1s_{1/2}$, $4f_{5/2}$-$1s_{1/2}$, $4f_{7/2}$-$1s_{1/2}$, $3p_{1/2}$-$2s_{1/2}$, $3p_{3/2}$-$2s_{1/2}$, $4d_{3/2}$-$2s_{1/2}$, $4d_{5/2}$-$2s_{1/2}$, $5f_{5/2}$-$2s_{1/2}$, and $5f_{7/2}$-$2s_{1/2}$.

It is a technical problem with respect to the numerical computation for searching the best-fit FB coefficients.
The dependence of Dirac energies on the FB coefficients can be deduced by the first-order perturbation theory as
\begin{equation}\label{eq-E-a}
	\delta E\sim\langle \delta V\rangle\sim\frac{\delta a_\nu}{\nu^2}.
\end{equation}
As a result, the parabolic behavior of  mean-square deviation  $\chi^2$ versus the FB coefficients is approximately satisfied,
\begin{equation}
	\chi^2\propto (a_\nu-a_\nu^{\min})^2,
\end{equation}
where $a_\nu^{\min}$ refer to the zero of the first-order partial derivative of $\chi^2$ with respect to $a_\nu$. 
Thus, the minimum value of $\chi^2$ for a certain coefficient can be easily converged by inverse parabolic interpolation~\cite{press2007numerical}.
Meanwhile, 
it is indicated from Eq.~(\ref{eq-E-a}) that the sensitivities of transitions to the coefficients satisfy the relationship: $a_1>a_2>\cdots>a_{N_{\mathrm{FB}}}$.  
A feasible iteration scheme is then established to converge $\chi^2$ to the minimum point within the $N_{\rm FB}$-dimension space, namely adjusting coefficients $a_\nu$ separately into $a_\nu^{\min}$ by inverse parabolic interpolation in order with the index $\nu$ from $1$ to $N_{\rm FB}-1$ at each iteration. 
When the coefficients are changed, the last coefficient $a_{N_{\rm FB}}$ is adjusted by Eq.~(\ref{eq-norm}) to keep the normalization. The initial set of coefficients (denoted by ``FB(init.)") are determined by fitting the 2pf distribution with parameters $t=2.1821$ fm and $c=6.6977$ fm which is the best fit under the constrain of transition energies (see our previous letter for details~\cite{xie2023revisiting}).


\begin{figure}
	\centering
	\includegraphics[width=8.5cm]{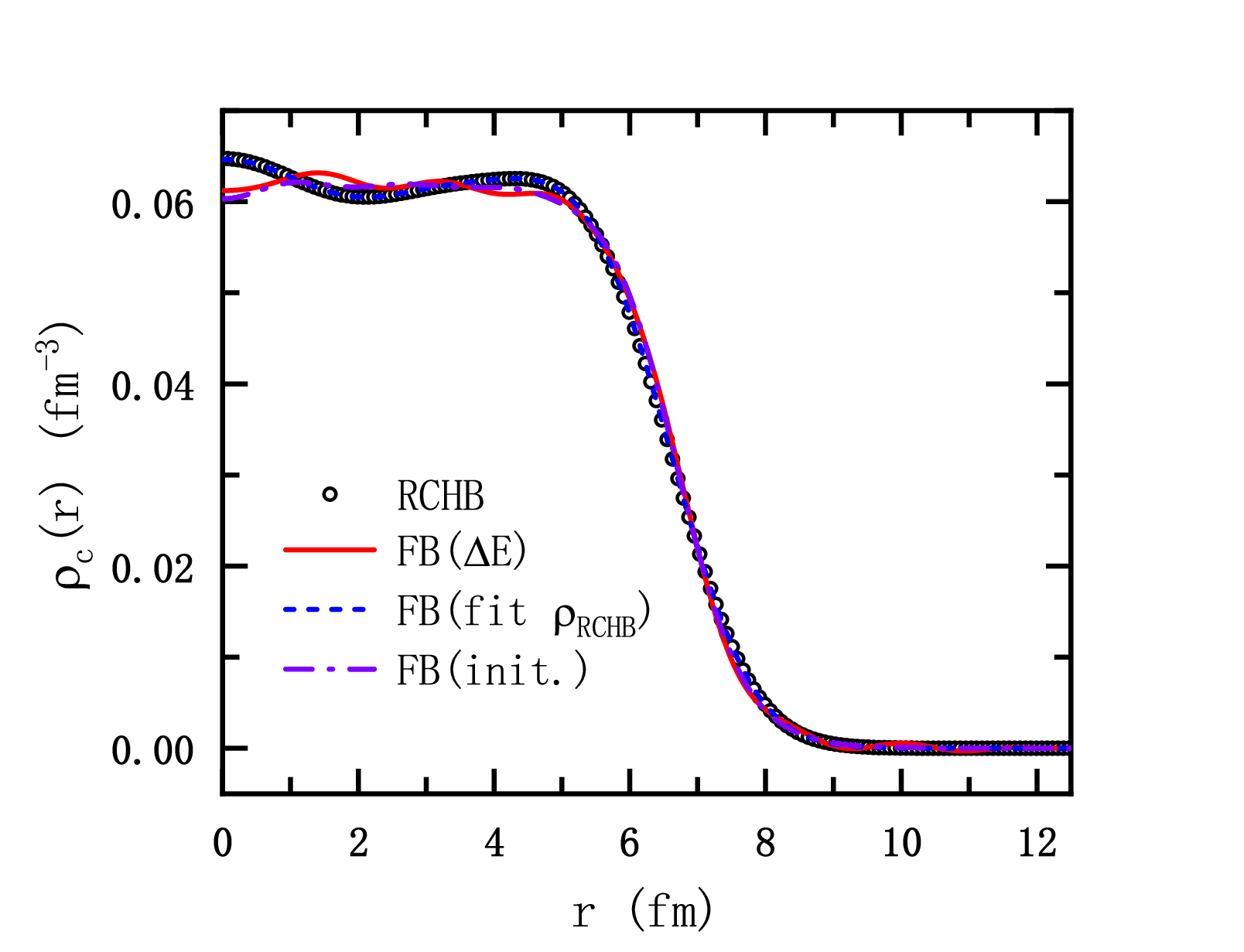}
	\caption{Comparison of nuclear charge densities between the RCHB calculations and there FB distributions for $^{208}$Pb. See the text for details.
	}
	\label{fig:fig1}
\end{figure}

\begin{table}
	\caption{The second, fourth, and sixth moments of the RCHB charge density for $^{208}$Pb, and the differences between the results of three FB distributions and those of the RCHB charge density.  The RMSD $\chi$ for transition energies and the deviation factor $\epsilon$ between three FB distributions and the RCHB charge density are also shown. See texts for details.}\label{tab-1}
	\centering
	\begin{tabular}{c|c|ccc}
		\toprule
		      &\multirow{2}{*}{RCHB}  &\multicolumn{3}{c}{The difference: FB-RCHB}   \\ &&FB(fit $\rho_\mathrm{RCHB}$)  &FB(init.)  &FB($\Delta E$) \\ \hline
  $R_2$(fm$^2$) &$30.310279$ &$-5.7\times10^{-5}$                &$0.012139$      &$-3\times10^{-6}$  \\
$R_4$(fm$^4$)   &$1166.040$  &$0.021$                            &$3.306$         &$0.025$            \\
$R_6$(fm$^6$)   &$51965.159$ &$6.705$                            &$671.486$       &$-1.769$           \\
$\chi$(keV)     &            &$1.3\times10^{-4}$                 &$0.34$          &$2.2\times10^{-5}$ \\
$\epsilon$      &            &$2.91\times10^{-4}$                &$0.030$         &$0.043$            \\
		\toprule
	\end{tabular}
\end{table}

\begin{figure}
	\centering
	\includegraphics[width=8.5cm]{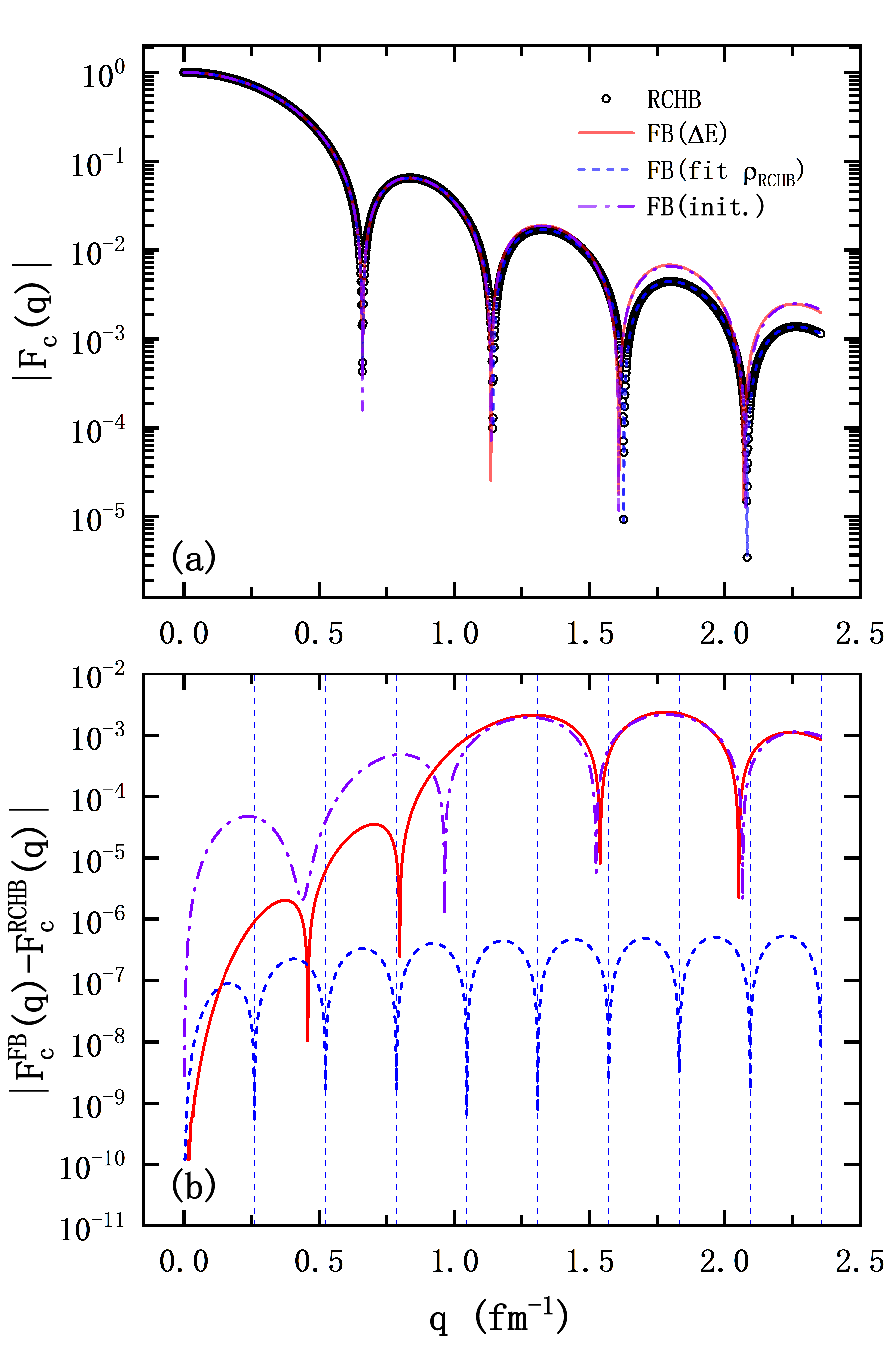}
	\caption{Charge form factors in $^{208}$Pb, where panel (a) shows the absolute value of charge FFs and panel (b) shows the absolute differences of charge FFs between three FB distributions and the targeted results. The vertical dotted lines denote the sampling points $q_\nu=\frac{\nu\pi}{R_\mathrm{cut}}$ with $\nu=1,2,3,\cdots$ 
	 }
	\label{fig:fig2}
\end{figure}

After performing the above iteration scheme, the RMSD $\chi$ is converged to $2.2\times 10^{-5}$ keV and the corresponding best-fit FB distribution, i.e., FB($\Delta E$), is compared with the RCHB charge density as shown in Table~\ref{tab-1} and Fig.~\ref{fig:fig1}. 
The results of FB distribution denoted by ``FB(fit $\rho_\mathrm{RCHB}$)" with coefficients determined by fitting the RCHB charge density through Eqs.~(\ref{eq-Fc-rhoc}) and (\ref{eq-a-Fc}) and the results of the initial set FB(init.) are also shown for comparison.
In Table~\ref{tab-1}, the differences of the second, fourth, and sixth moments between three FB distributions and the RCHB charge density, as well as the deviation factor $\epsilon$ and the RMSD $\chi$ for transition energies are displayed. 
It can be seen that the second, fourth, and sixth moments of the best-fit FB distribution obtained from the above iteration scheme are in an excellent agreement with the benchmark, even better than that of the FB(fit $\rho_\mathrm{RCHB}$) fitting the density distribution directly. 
This confirms the possibility of extracting the second, fourth, and sixth moments of nuclear charge density from the muonic atom spectroscopy. 
However, the deviation factor $\epsilon$ between them is not smaller but bigger than the initial value of $0.03$. 
Meanwhile, the RMSD $\chi$ for transition energies for the FB(fit $\rho_\mathrm{RCHB}$) is an order of magnitude larger than that of the best-fit FB distribution. 
This indicates that FB($\Delta E$) obtained under the constraint of transitions with a finite number of FB coefficients would not converge to an FB distribution that closes to the targeted density.

In Fig.~\ref{fig:fig1}, three FB distributions are compared with the RCHB charge density. 
It can be visibly seen that the best-fit FB distribution does not reproduce the targeted density in detail. 
This indicates that the transitions are not sensitive to the detailed charge density. 
Thus, one can conclude that extracting detailed nuclear charge density from the muonic atom spectroscopy is not practical, but the extractions of the second, fourth, and sixth moments with high accuracy are possible.

\begin{figure}
	\centering
	\includegraphics[width=8.5cm]{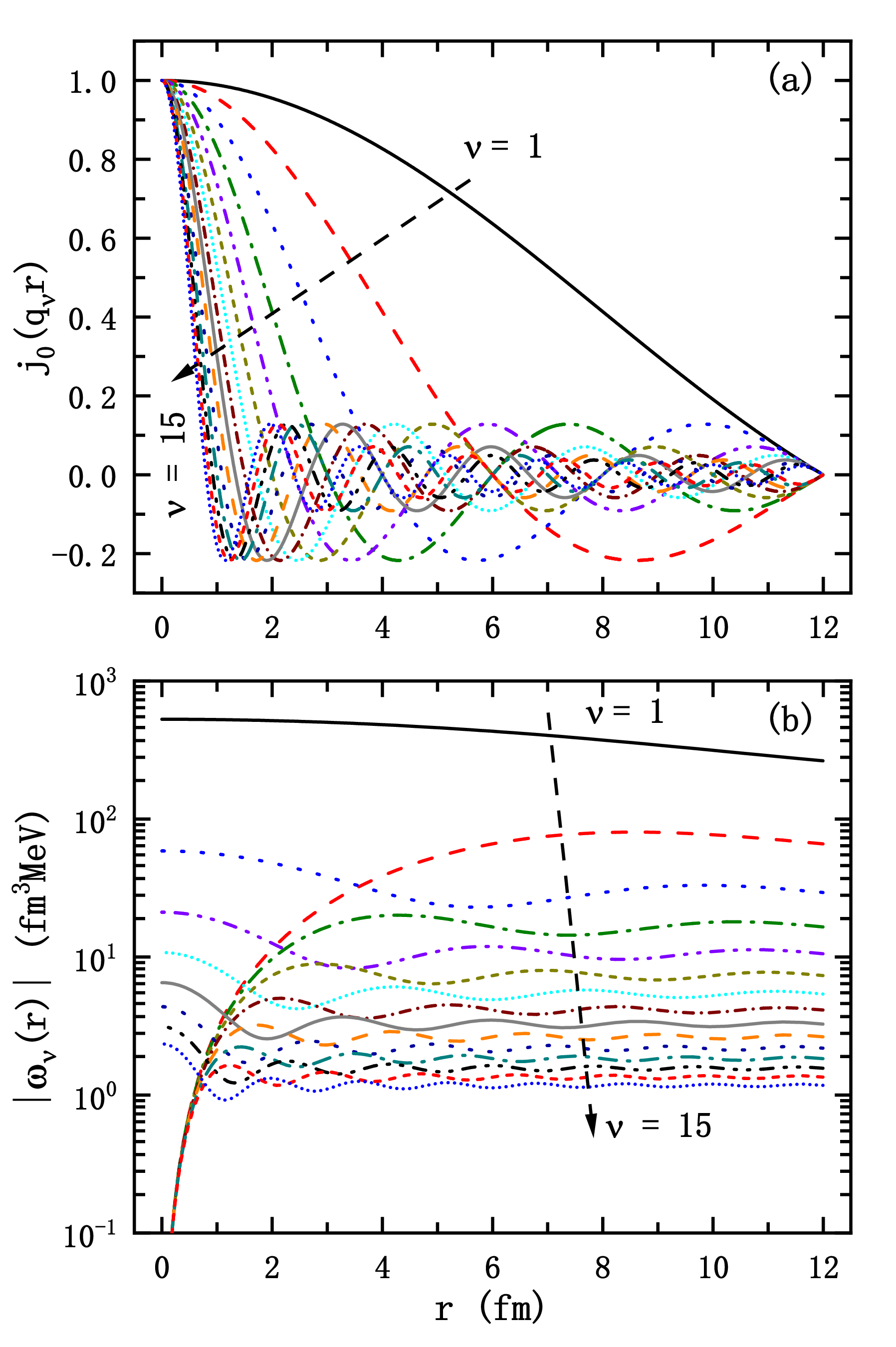}
	\caption{The weight functions with respect to the charge density and electric potential. Panel (a) shows the zeroth-order spherical Bessel functions related to the FB coefficients and panel (b) shows the absolute value of the weight functions of FB coefficients for the electrostatic potential given by Eq.~(\ref{eq-omega}). 
	}
	\label{fig:fig3}
\end{figure}

The charge FFs are also of interest here considering the relationship between charge FF $F_c$ and the $2k$-th moments $R_{2k}$ derived from Eq.~(\ref{eq-Fc-rhoc}) as
\begin{equation}
	F_c(q)=Z\sum_{k=0}^\infty\frac{R_{2k}q^{2k}}{(2k+1)!}
\end{equation}
and 
\begin{equation}
	R_{2k}=-\frac{(2k+1)!}{k!}\frac{\partial^kF_c(q)}{\partial(q^2)^k}.
\end{equation}
The charge FFs of three FB distributions are compared in Fig.~\ref{fig:fig2} with the benchmark, i.e., the charge FFs corresponding to the RCHB charge density. 
It can be seen from Fig.~\ref{fig:fig2}(a) that the charge FFs of FB(init.) and FB($\Delta E$) gradually deviate from the benchmark with increasing $q$, while the charge FF of FB(fit $\rho_{\rm RCHB}$) is in an excellent agreement with the benchmark. 
Furthermore, the absolute differences of charge FFs between three FB distributions and the benchmark are shown in Fig.~\ref{fig:fig2}(b). One can see that the charge FF of the best fit is in a better agreement with the benchmark than that of FB(fit $\rho_{\rm RCHB}$) in the low-$q$ region ($q<0.2~\mathrm{fm}^{-1}$). 
Therefore, it can be concluded that the transitions are quite sensitive to the low-$q$ region of charge FF, i.e., it is possible to extract charge FF in low-$q$ region from the muonic atom spectroscopy.

\begin{figure}
	\centering
	\includegraphics[width=8.5cm]{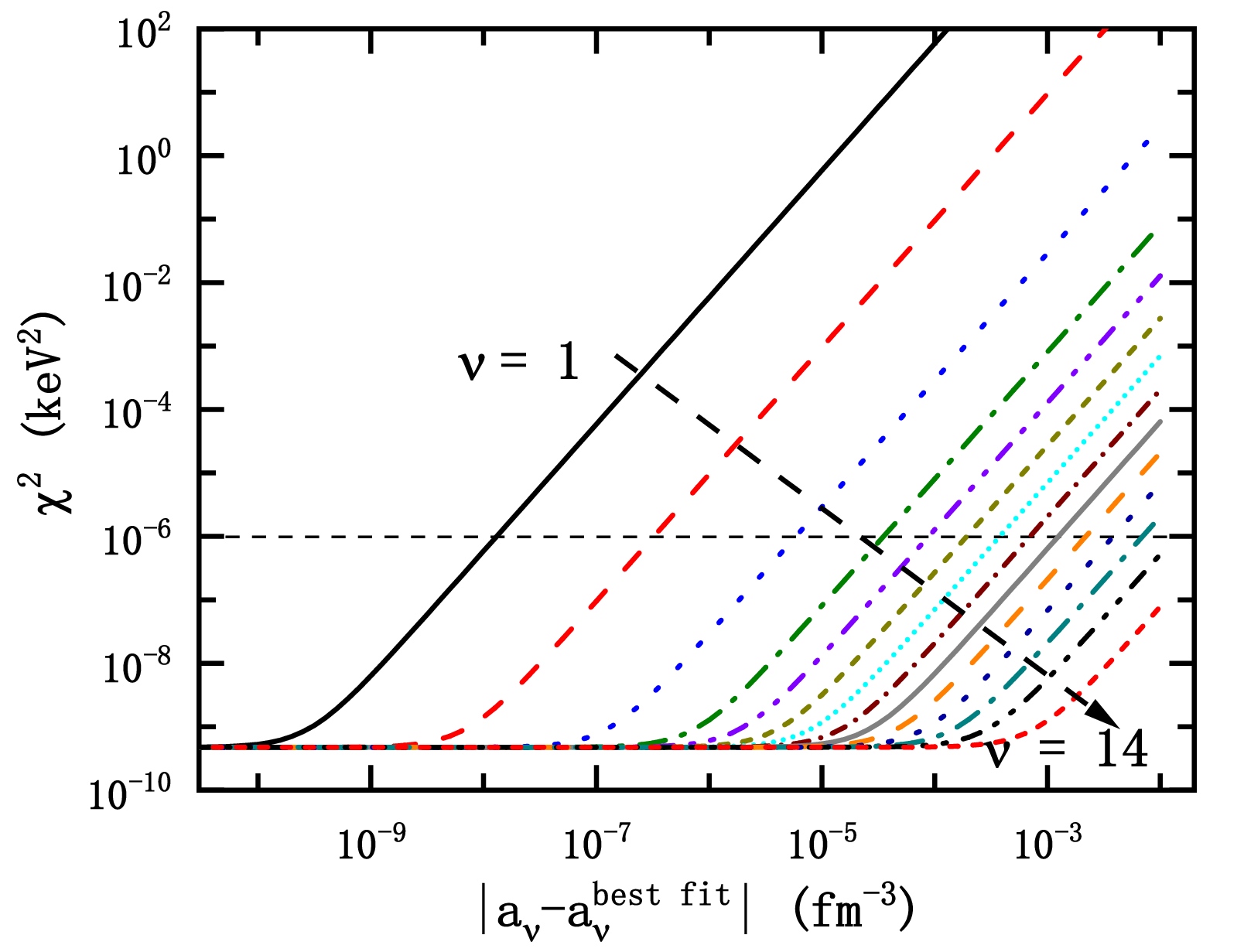}
	\caption{The mean-square deviation $\chi^2$ versus the deviation of FB coeffcients from the best fit $a_\nu^\mathrm{best~fit}$ with the index of coefficient from $1$ to $14$. The last coefficient $a_{15}$ is adjusted to maintain normalization of Eq.~(\ref{eq-norm}). 
	}
	\label{fig:fig4}
\end{figure}

On the other hand, according to Eq.~(\ref{eq-a-Fc}), the $\nu$-th coefficient $a_\nu$ is directly related to the charge FF $F_c(q)$ at $q=q_\nu$. 
To better understand the results of Figs.~\ref{fig:fig1} and~\ref{fig:fig2}, the sensitivities of charge density, electrostatic potential, and transition energies to the FB coefficients are investigated and indicated in Figs.~\ref{fig:fig3} and~\ref{fig:fig4}. Figure~\ref{fig:fig3} shows the weight functions of the FB coefficients $a_\nu$ with $\nu=1,2,\cdots,15$ for the charge density and electrostatic potential, i.e., $j_0(q_\nu r)$ and $\omega_\nu(r)$. 
It can be seen from Fig.~\ref{fig:fig3}(a) that the weights are identical as $j_0(q_\nu r)=1$ for all coefficients at the center $r=0$. 
As a result, all of these coefficients are of almost equal importance for the description of charge density in the center region. 
In comparison, the leading-order coefficients are visibly more important than the higher-order coefficients for the electrostatic potential as shown in Fig.~\ref{fig:fig3}(b). 
Thus, it is considered that the leading-order coefficients should also be more important for the transitions from the perspective of perturbation theory.

In Fig.~\ref{fig:fig4}, the variations of mean-square deviation $\chi^2$ with the FB coefficients deviating from the best fit are shown. 
It is indicated that the transitions are extremely sensitive to the leading-order coefficients and are insensitive to the higher-order coefficients. 
For example, a deviation of $a_1$ from the best fit by only $10^{-8}$ fm$^{-3}$ makes $\chi^2$ increased to $10^{-6}$ keV$^2$, in comparison with the deviation of $a_{12}$ from the best fit by $0.008$ fm$^{-3}$. 
As a result, the great arbitrariness of higher-order coefficients makes it difficult to extract the detailed charge density from the muonic atom spectroscopy, and it can be seen from Fig~\ref{fig:fig2}(b) that the charge FF of FB($\Delta E$) tends to that of the initial distribution (namely FB(init.)) with increasing $q$, since the higher-order coefficients cannot be constrained well by the transitions.

It should be noted that the shape and charge FF (for $q>0.2$ fm$^{-1}$) of the best-fit FB distribution shown in Figs.~\ref{fig:fig1} and \ref{fig:fig2} depend on the initial set FB(init.). 
For example, if FB(fit $\rho_\mathrm{RCHB}$) is taken as the initial set, the results of the best fit converged by the present iteration scheme would be closer to the benchmark, for the reason that the sensitivities of transitions to the coefficients $a_\nu$ are weaker with the increasing index $\nu$. 
While the second, fourth, and sixth moments as well as the charge FF at low-$q$ can be invariably extracted from the muonic atom spectroscopy whatever the initial FB distribution is taken as discussed above. 
However, such a procedure, i.e., taking FB(fit $\rho_\mathrm{RCHB}$) as the initial set, is not feasible in practice, since the transition energies are supposed to be the only available information when the corresponding analysis is carried out.

\section{Summary}\label{sec4}

In this work, it is explored how much information can be extracted from the muonic atom spectroscopy using a model-independent density distribution, i.e., the FB expansion. 
Taking $^{208}$Pb as an example and taking the RCHB charge density as the benchmark, an iteration scheme has been made and performed to search the minimum value of the RMSD $\chi$ and the corresponding best-fit FB coefficients. 
By comparing the second, fourth, and sixth moments of three FB distributions, i.e., FB($\Delta E$), FB(fit $\rho_{\rm RCHB}$), and FB(init.) with the targeted values, it is concluded that the extractions of the second, fourth, and sixth moments can be performed with high accuracy. Furthermore, the charge densities and charge FF of three FB distributions are compared with the benchmark. 
It is shown that the best fit cannot reproduce the detailed charge density but can well describe the charge FF in the low-$q$ region. 
In the end, the sensitivities of charge density, electrostatic potential, and transitions to the FB coefficients are investigated and indicated that the difficulty of the extraction of detailed charge density originates from the arbitrariness of higher-order FB coefficients, since the transitions are insensitive to them.

\begin{acknowledgments}
We would like to thank Dr.~Tomoya Naito for the fruitful discussions.  
This work was supported by
the National Natural Science Foundation of China (No.~11675063), the Natural Science Foundation of Jilin Province (No.~20220101017JC),
the Key Laboratory of Nuclear Data Foundation (JCKY2020201C157),
the JSPS Grant-in-Aid for Scientific Research (S) under Grant No.~20H05648,
the RIKEN iTHEMS Program, 
and the RIKEN Pioneering Project: Evolution of Matter in the Universe.
\end{acknowledgments}

\nocite{*}

\bibliography{apssamp}

\providecommand{\noopsort}[1]{}\providecommand{\singleletter}[1]{#1}%
\begin{thebibliography}{71}%
\makeatletter
\providecommand \@ifxundefined [1]{%
 \@ifx{#1\undefined}
}%
\providecommand \@ifnum [1]{%
 \ifnum #1\expandafter \@firstoftwo
 \else \expandafter \@secondoftwo
 \fi
}%
\providecommand \@ifx [1]{%
 \ifx #1\expandafter \@firstoftwo
 \else \expandafter \@secondoftwo
 \fi
}%
\providecommand \natexlab [1]{#1}%
\providecommand \enquote  [1]{``#1''}%
\providecommand \bibnamefont  [1]{#1}%
\providecommand \bibfnamefont [1]{#1}%
\providecommand \citenamefont [1]{#1}%
\providecommand \href@noop [0]{\@secondoftwo}%
\providecommand \href [0]{\begingroup \@sanitize@url \@href}%
\providecommand \@href[1]{\@@startlink{#1}\@@href}%
\providecommand \@@href[1]{\endgroup#1\@@endlink}%
\providecommand \@sanitize@url [0]{\catcode `\\12\catcode `\$12\catcode `\&12\catcode `\#12\catcode `\^12\catcode `\_12\catcode `\%12\relax}%
\providecommand \@@startlink[1]{}%
\providecommand \@@endlink[0]{}%
\providecommand \url  [0]{\begingroup\@sanitize@url \@url }%
\providecommand \@url [1]{\endgroup\@href {#1}{\urlprefix }}%
\providecommand \urlprefix  [0]{URL }%
\providecommand \Eprint [0]{\href }%
\providecommand \doibase [0]{https://doi.org/}%
\providecommand \selectlanguage [0]{\@gobble}%
\providecommand \bibinfo  [0]{\@secondoftwo}%
\providecommand \bibfield  [0]{\@secondoftwo}%
\providecommand \translation [1]{[#1]}%
\providecommand \BibitemOpen [0]{}%
\providecommand \bibitemStop [0]{}%
\providecommand \bibitemNoStop [0]{.\EOS\space}%
\providecommand \EOS [0]{\spacefactor3000\relax}%
\providecommand \BibitemShut  [1]{\csname bibitem#1\endcsname}%
\let\auto@bib@innerbib\@empty
\bibitem [{\citenamefont {Angeli}(2004)}]{ANGELI2004185}%
  \BibitemOpen
  \bibfield  {author} {\bibinfo {author} {\bibfnamefont {I.}~\bibnamefont {Angeli}},\ }\href {https://doi.org/https://doi.org/10.1016/j.adt.2004.04.002} {\bibfield  {journal} {\bibinfo  {journal} {At. Data Nucl. Data Tables}\ }\textbf {\bibinfo {volume} {87}},\ \bibinfo {pages} {185} (\bibinfo {year} {2004})}\BibitemShut {NoStop}%
\bibitem [{\citenamefont {Angeli}\ and\ \citenamefont {Marinova}(2013)}]{ANGELI201369}%
  \BibitemOpen
  \bibfield  {author} {\bibinfo {author} {\bibfnamefont {I.}~\bibnamefont {Angeli}}\ and\ \bibinfo {author} {\bibfnamefont {K.}~\bibnamefont {Marinova}},\ }\href {https://doi.org/https://doi.org/10.1016/j.adt.2011.12.006} {\bibfield  {journal} {\bibinfo  {journal} {At. Data Nucl. Data Tables}\ }\textbf {\bibinfo {volume} {99}},\ \bibinfo {pages} {69} (\bibinfo {year} {2013})}\BibitemShut {NoStop}%
\bibitem [{\citenamefont {Li}\ \emph {et~al.}(2021)\citenamefont {Li}, \citenamefont {Luo},\ and\ \citenamefont {Wang}}]{LI2021101440}%
  \BibitemOpen
  \bibfield  {author} {\bibinfo {author} {\bibfnamefont {T.}~\bibnamefont {Li}}, \bibinfo {author} {\bibfnamefont {Y.}~\bibnamefont {Luo}},\ and\ \bibinfo {author} {\bibfnamefont {N.}~\bibnamefont {Wang}},\ }\href {https://doi.org/https://doi.org/10.1016/j.adt.2021.101440} {\bibfield  {journal} {\bibinfo  {journal} {At. Data Nucl. Data Tables}\ }\textbf {\bibinfo {volume} {140}},\ \bibinfo {pages} {101440} (\bibinfo {year} {2021})}\BibitemShut {NoStop}%
\bibitem [{\citenamefont {Hofstadter}(1956)}]{RevModPhys.28.214}%
  \BibitemOpen
  \bibfield  {author} {\bibinfo {author} {\bibfnamefont {R.}~\bibnamefont {Hofstadter}},\ }\href {https://doi.org/10.1103/RevModPhys.28.214} {\bibfield  {journal} {\bibinfo  {journal} {Rev. Mod. Phys.}\ }\textbf {\bibinfo {volume} {28}},\ \bibinfo {pages} {214} (\bibinfo {year} {1956})}\BibitemShut {NoStop}%
\bibitem [{\citenamefont {Dreher}\ \emph {et~al.}(1974)\citenamefont {Dreher}, \citenamefont {Friedrich}, \citenamefont {Merle}, \citenamefont {Rothhaas},\ and\ \citenamefont {Lührs}}]{DREHER1974219}%
  \BibitemOpen
  \bibfield  {author} {\bibinfo {author} {\bibfnamefont {B.}~\bibnamefont {Dreher}}, \bibinfo {author} {\bibfnamefont {J.}~\bibnamefont {Friedrich}}, \bibinfo {author} {\bibfnamefont {K.}~\bibnamefont {Merle}}, \bibinfo {author} {\bibfnamefont {H.}~\bibnamefont {Rothhaas}},\ and\ \bibinfo {author} {\bibfnamefont {G.}~\bibnamefont {Lührs}},\ }\href {https://doi.org/https://doi.org/10.1016/0375-9474(74)90189-4} {\bibfield  {journal} {\bibinfo  {journal} {Nucl. Phys. A}\ }\textbf {\bibinfo {volume} {235}},\ \bibinfo {pages} {219} (\bibinfo {year} {1974})}\BibitemShut {NoStop}%
\bibitem [{\citenamefont {{De Vries}}\ \emph {et~al.}(1987)\citenamefont {{De Vries}}, \citenamefont {{De Jager}},\ and\ \citenamefont {{De Vries}}}]{DEVRIES1987495}%
  \BibitemOpen
  \bibfield  {author} {\bibinfo {author} {\bibfnamefont {H.}~\bibnamefont {{De Vries}}}, \bibinfo {author} {\bibfnamefont {C.}~\bibnamefont {{De Jager}}},\ and\ \bibinfo {author} {\bibfnamefont {C.}~\bibnamefont {{De Vries}}},\ }\href {https://doi.org/https://doi.org/10.1016/0092-640X(87)90013-1} {\bibfield  {journal} {\bibinfo  {journal} {At. Data Nucl. Data Tables}\ }\textbf {\bibinfo {volume} {36}},\ \bibinfo {pages} {495} (\bibinfo {year} {1987})}\BibitemShut {NoStop}%
\bibitem [{\citenamefont {Engfer}\ \emph {et~al.}(1974)\citenamefont {Engfer}, \citenamefont {Schneuwly}, \citenamefont {Vuilleumier}, \citenamefont {Walter},\ and\ \citenamefont {Zehnder}}]{ENGFER1974509}%
  \BibitemOpen
  \bibfield  {author} {\bibinfo {author} {\bibfnamefont {R.}~\bibnamefont {Engfer}}, \bibinfo {author} {\bibfnamefont {H.}~\bibnamefont {Schneuwly}}, \bibinfo {author} {\bibfnamefont {J.}~\bibnamefont {Vuilleumier}}, \bibinfo {author} {\bibfnamefont {H.}~\bibnamefont {Walter}},\ and\ \bibinfo {author} {\bibfnamefont {A.}~\bibnamefont {Zehnder}},\ }\href {https://doi.org/https://doi.org/10.1016/S0092-640X(74)80003-3} {\bibfield  {journal} {\bibinfo  {journal} {At. Data Nucl. Data Tables}\ }\textbf {\bibinfo {volume} {14}},\ \bibinfo {pages} {509} (\bibinfo {year} {1974})}\BibitemShut {NoStop}%
\bibitem [{\citenamefont {Wohlfahrt}\ \emph {et~al.}(1981)\citenamefont {Wohlfahrt}, \citenamefont {Shera}, \citenamefont {Hoehn}, \citenamefont {Yamazaki},\ and\ \citenamefont {Steffen}}]{PhysRevC.23.533}%
  \BibitemOpen
  \bibfield  {author} {\bibinfo {author} {\bibfnamefont {H.~D.}\ \bibnamefont {Wohlfahrt}}, \bibinfo {author} {\bibfnamefont {E.~B.}\ \bibnamefont {Shera}}, \bibinfo {author} {\bibfnamefont {M.~V.}\ \bibnamefont {Hoehn}}, \bibinfo {author} {\bibfnamefont {Y.}~\bibnamefont {Yamazaki}},\ and\ \bibinfo {author} {\bibfnamefont {R.~M.}\ \bibnamefont {Steffen}},\ }\href {https://doi.org/10.1103/PhysRevC.23.533} {\bibfield  {journal} {\bibinfo  {journal} {Phys. Rev. C}\ }\textbf {\bibinfo {volume} {23}},\ \bibinfo {pages} {533} (\bibinfo {year} {1981})}\BibitemShut {NoStop}%
\bibitem [{\citenamefont {Bergem}\ \emph {et~al.}(1988)\citenamefont {Bergem}, \citenamefont {Piller}, \citenamefont {Rueetschi}, \citenamefont {Schaller}, \citenamefont {Schellenberg},\ and\ \citenamefont {Schneuwly}}]{PhysRevC.37.2821}%
  \BibitemOpen
  \bibfield  {author} {\bibinfo {author} {\bibfnamefont {P.}~\bibnamefont {Bergem}}, \bibinfo {author} {\bibfnamefont {G.}~\bibnamefont {Piller}}, \bibinfo {author} {\bibfnamefont {A.}~\bibnamefont {Rueetschi}}, \bibinfo {author} {\bibfnamefont {L.~A.}\ \bibnamefont {Schaller}}, \bibinfo {author} {\bibfnamefont {L.}~\bibnamefont {Schellenberg}},\ and\ \bibinfo {author} {\bibfnamefont {H.}~\bibnamefont {Schneuwly}},\ }\href {https://doi.org/10.1103/PhysRevC.37.2821} {\bibfield  {journal} {\bibinfo  {journal} {Phys. Rev. C}\ }\textbf {\bibinfo {volume} {37}},\ \bibinfo {pages} {2821} (\bibinfo {year} {1988})}\BibitemShut {NoStop}%
\bibitem [{\citenamefont {Schaller}\ \emph {et~al.}(1978)\citenamefont {Schaller}, \citenamefont {Dubler}, \citenamefont {Kaeser}, \citenamefont {Rinker}, \citenamefont {Robert-Tissot}, \citenamefont {Schellenberg},\ and\ \citenamefont {Schneuwly}}]{SCHALLER1978225}%
  \BibitemOpen
  \bibfield  {author} {\bibinfo {author} {\bibfnamefont {L.}~\bibnamefont {Schaller}}, \bibinfo {author} {\bibfnamefont {T.}~\bibnamefont {Dubler}}, \bibinfo {author} {\bibfnamefont {K.}~\bibnamefont {Kaeser}}, \bibinfo {author} {\bibfnamefont {G.}~\bibnamefont {Rinker}}, \bibinfo {author} {\bibfnamefont {B.}~\bibnamefont {Robert-Tissot}}, \bibinfo {author} {\bibfnamefont {L.}~\bibnamefont {Schellenberg}},\ and\ \bibinfo {author} {\bibfnamefont {H.}~\bibnamefont {Schneuwly}},\ }\href {https://doi.org/https://doi.org/10.1016/0375-9474(78)96128-6} {\bibfield  {journal} {\bibinfo  {journal} {Nucl. Phys. A}\ }\textbf {\bibinfo {volume} {300}},\ \bibinfo {pages} {225} (\bibinfo {year} {1978})}\BibitemShut {NoStop}%
\bibitem [{\citenamefont {Yamazaki}\ \emph {et~al.}(1978)\citenamefont {Yamazaki}, \citenamefont {Shera}, \citenamefont {Hoehn},\ and\ \citenamefont {Steffen}}]{PhysRevC.18.1474}%
  \BibitemOpen
  \bibfield  {author} {\bibinfo {author} {\bibfnamefont {Y.}~\bibnamefont {Yamazaki}}, \bibinfo {author} {\bibfnamefont {E.~B.}\ \bibnamefont {Shera}}, \bibinfo {author} {\bibfnamefont {M.~V.}\ \bibnamefont {Hoehn}},\ and\ \bibinfo {author} {\bibfnamefont {R.~M.}\ \bibnamefont {Steffen}},\ }\href {https://doi.org/10.1103/PhysRevC.18.1474} {\bibfield  {journal} {\bibinfo  {journal} {Phys. Rev. C}\ }\textbf {\bibinfo {volume} {18}},\ \bibinfo {pages} {1474} (\bibinfo {year} {1978})}\BibitemShut {NoStop}%
\bibitem [{\citenamefont {Shera}\ \emph {et~al.}(1989)\citenamefont {Shera}, \citenamefont {Hoehn}, \citenamefont {Fricke},\ and\ \citenamefont {Mallot}}]{PhysRevC.39.195}%
  \BibitemOpen
  \bibfield  {author} {\bibinfo {author} {\bibfnamefont {E.~B.}\ \bibnamefont {Shera}}, \bibinfo {author} {\bibfnamefont {M.~V.}\ \bibnamefont {Hoehn}}, \bibinfo {author} {\bibfnamefont {G.}~\bibnamefont {Fricke}},\ and\ \bibinfo {author} {\bibfnamefont {G.}~\bibnamefont {Mallot}},\ }\href {https://doi.org/10.1103/PhysRevC.39.195} {\bibfield  {journal} {\bibinfo  {journal} {Phys. Rev. C}\ }\textbf {\bibinfo {volume} {39}},\ \bibinfo {pages} {195} (\bibinfo {year} {1989})}\BibitemShut {NoStop}%
\bibitem [{\citenamefont {Knecht}\ \emph {et~al.}(2020)\citenamefont {Knecht}, \citenamefont {Skawran},\ and\ \citenamefont {Vogiatzi}}]{Knecht2020}%
  \BibitemOpen
  \bibfield  {author} {\bibinfo {author} {\bibfnamefont {A.}~\bibnamefont {Knecht}}, \bibinfo {author} {\bibfnamefont {A.}~\bibnamefont {Skawran}},\ and\ \bibinfo {author} {\bibfnamefont {S.~M.}\ \bibnamefont {Vogiatzi}},\ }\href {https://doi.org/10.1140/epjp/s13360-020-00777-y} {\bibfield  {journal} {\bibinfo  {journal} {Eur. Phys. J. Plus}\ }\textbf {\bibinfo {volume} {135}},\ \bibinfo {pages} {777} (\bibinfo {year} {2020})}\BibitemShut {NoStop}%
\bibitem [{\citenamefont {Campbell}\ \emph {et~al.}(2016)\citenamefont {Campbell}, \citenamefont {Moore},\ and\ \citenamefont {Pearson}}]{CAMPBELL2016127}%
  \BibitemOpen
  \bibfield  {author} {\bibinfo {author} {\bibfnamefont {P.}~\bibnamefont {Campbell}}, \bibinfo {author} {\bibfnamefont {I.}~\bibnamefont {Moore}},\ and\ \bibinfo {author} {\bibfnamefont {M.}~\bibnamefont {Pearson}},\ }\href {https://doi.org/https://doi.org/10.1016/j.ppnp.2015.09.003} {\bibfield  {journal} {\bibinfo  {journal} {Prog. Part. Nucl. Phys.}\ }\textbf {\bibinfo {volume} {86}},\ \bibinfo {pages} {127} (\bibinfo {year} {2016})}\BibitemShut {NoStop}%
\bibitem [{\citenamefont {Heylen}\ \emph {et~al.}(2016)\citenamefont {Heylen}, \citenamefont {Babcock}, \citenamefont {Beerwerth}, \citenamefont {Billowes}, \citenamefont {Bissell}, \citenamefont {Blaum}, \citenamefont {Bonnard}, \citenamefont {Campbell}, \citenamefont {Cheal}, \citenamefont {Day~Goodacre}, \citenamefont {Fedorov}, \citenamefont {Fritzsche}, \citenamefont {Garcia~Ruiz}, \citenamefont {Geithner}, \citenamefont {Geppert}, \citenamefont {Gins}, \citenamefont {Grob}, \citenamefont {Kowalska}, \citenamefont {Kreim}, \citenamefont {Lenzi}, \citenamefont {Moore}, \citenamefont {Maass}, \citenamefont {Malbrunot-Ettenauer}, \citenamefont {Marsh}, \citenamefont {Neugart}, \citenamefont {Neyens}, \citenamefont {N\"ortersh\"auser}, \citenamefont {Otsuka}, \citenamefont {Papuga}, \citenamefont {Rossel}, \citenamefont {Rothe}, \citenamefont {S\'anchez}, \citenamefont {Tsunoda}, \citenamefont {Wraith}, \citenamefont {Xie}, \citenamefont {Yang},\ and\ \citenamefont {Yordanov}}]{PhysRevC.94.054321}%
  \BibitemOpen
  \bibfield  {author} {\bibinfo {author} {\bibfnamefont {H.}~\bibnamefont {Heylen}}, \bibinfo {author} {\bibfnamefont {C.}~\bibnamefont {Babcock}}, \bibinfo {author} {\bibfnamefont {R.}~\bibnamefont {Beerwerth}}, \bibinfo {author} {\bibfnamefont {J.}~\bibnamefont {Billowes}}, \bibinfo {author} {\bibfnamefont {M.~L.}\ \bibnamefont {Bissell}}, \bibinfo {author} {\bibfnamefont {K.}~\bibnamefont {Blaum}}, \bibinfo {author} {\bibfnamefont {J.}~\bibnamefont {Bonnard}}, \bibinfo {author} {\bibfnamefont {P.}~\bibnamefont {Campbell}}, \bibinfo {author} {\bibfnamefont {B.}~\bibnamefont {Cheal}}, \bibinfo {author} {\bibfnamefont {T.}~\bibnamefont {Day~Goodacre}}, \bibinfo {author} {\bibfnamefont {D.}~\bibnamefont {Fedorov}}, \bibinfo {author} {\bibfnamefont {S.}~\bibnamefont {Fritzsche}}, \bibinfo {author} {\bibfnamefont {R.~F.}\ \bibnamefont {Garcia~Ruiz}}, \bibinfo {author} {\bibfnamefont {W.}~\bibnamefont {Geithner}}, \bibinfo {author} {\bibfnamefont {C.}~\bibnamefont {Geppert}}, \bibinfo {author} {\bibfnamefont
  {W.}~\bibnamefont {Gins}}, \bibinfo {author} {\bibfnamefont {L.~K.}\ \bibnamefont {Grob}}, \bibinfo {author} {\bibfnamefont {M.}~\bibnamefont {Kowalska}}, \bibinfo {author} {\bibfnamefont {K.}~\bibnamefont {Kreim}}, \bibinfo {author} {\bibfnamefont {S.~M.}\ \bibnamefont {Lenzi}}, \bibinfo {author} {\bibfnamefont {I.~D.}\ \bibnamefont {Moore}}, \bibinfo {author} {\bibfnamefont {B.}~\bibnamefont {Maass}}, \bibinfo {author} {\bibfnamefont {S.}~\bibnamefont {Malbrunot-Ettenauer}}, \bibinfo {author} {\bibfnamefont {B.}~\bibnamefont {Marsh}}, \bibinfo {author} {\bibfnamefont {R.}~\bibnamefont {Neugart}}, \bibinfo {author} {\bibfnamefont {G.}~\bibnamefont {Neyens}}, \bibinfo {author} {\bibfnamefont {W.}~\bibnamefont {N\"ortersh\"auser}}, \bibinfo {author} {\bibfnamefont {T.}~\bibnamefont {Otsuka}}, \bibinfo {author} {\bibfnamefont {J.}~\bibnamefont {Papuga}}, \bibinfo {author} {\bibfnamefont {R.}~\bibnamefont {Rossel}}, \bibinfo {author} {\bibfnamefont {S.}~\bibnamefont {Rothe}}, \bibinfo {author} {\bibfnamefont
  {R.}~\bibnamefont {S\'anchez}}, \bibinfo {author} {\bibfnamefont {Y.}~\bibnamefont {Tsunoda}}, \bibinfo {author} {\bibfnamefont {C.}~\bibnamefont {Wraith}}, \bibinfo {author} {\bibfnamefont {L.}~\bibnamefont {Xie}}, \bibinfo {author} {\bibfnamefont {X.~F.}\ \bibnamefont {Yang}},\ and\ \bibinfo {author} {\bibfnamefont {D.~T.}\ \bibnamefont {Yordanov}},\ }\href {https://doi.org/10.1103/PhysRevC.94.054321} {\bibfield  {journal} {\bibinfo  {journal} {Phys. Rev. C}\ }\textbf {\bibinfo {volume} {94}},\ \bibinfo {pages} {054321} (\bibinfo {year} {2016})}\BibitemShut {NoStop}%
\bibitem [{\citenamefont {Koszor{\'u}s}\ \emph {et~al.}(2021)\citenamefont {Koszor{\'u}s}, \citenamefont {Yang}, \citenamefont {Jiang}, \citenamefont {Novario}, \citenamefont {Bai}, \citenamefont {Billowes}, \citenamefont {Binnersley}, \citenamefont {Bissell}, \citenamefont {Cocolios}, \citenamefont {Cooper}, \citenamefont {de~Groote}, \citenamefont {Ekstr{\"o}m}, \citenamefont {Flanagan}, \citenamefont {Forss{\'e}n}, \citenamefont {Franchoo}, \citenamefont {Ruiz}, \citenamefont {Gustafsson}, \citenamefont {Hagen}, \citenamefont {Jansen}, \citenamefont {Kanellakopoulos}, \citenamefont {Kortelainen}, \citenamefont {Nazarewicz}, \citenamefont {Neyens}, \citenamefont {Papenbrock}, \citenamefont {Reinhard}, \citenamefont {Ricketts}, \citenamefont {Sahoo}, \citenamefont {Vernon},\ and\ \citenamefont {Wilkins}}]{Koszorus2021}%
  \BibitemOpen
  \bibfield  {author} {\bibinfo {author} {\bibfnamefont {{\'A}.}~\bibnamefont {Koszor{\'u}s}}, \bibinfo {author} {\bibfnamefont {X.~F.}\ \bibnamefont {Yang}}, \bibinfo {author} {\bibfnamefont {W.~G.}\ \bibnamefont {Jiang}}, \bibinfo {author} {\bibfnamefont {S.~J.}\ \bibnamefont {Novario}}, \bibinfo {author} {\bibfnamefont {S.~W.}\ \bibnamefont {Bai}}, \bibinfo {author} {\bibfnamefont {J.}~\bibnamefont {Billowes}}, \bibinfo {author} {\bibfnamefont {C.~L.}\ \bibnamefont {Binnersley}}, \bibinfo {author} {\bibfnamefont {M.~L.}\ \bibnamefont {Bissell}}, \bibinfo {author} {\bibfnamefont {T.~E.}\ \bibnamefont {Cocolios}}, \bibinfo {author} {\bibfnamefont {B.~S.}\ \bibnamefont {Cooper}}, \bibinfo {author} {\bibfnamefont {R.~P.}\ \bibnamefont {de~Groote}}, \bibinfo {author} {\bibfnamefont {A.}~\bibnamefont {Ekstr{\"o}m}}, \bibinfo {author} {\bibfnamefont {K.~T.}\ \bibnamefont {Flanagan}}, \bibinfo {author} {\bibfnamefont {C.}~\bibnamefont {Forss{\'e}n}}, \bibinfo {author} {\bibfnamefont {S.}~\bibnamefont {Franchoo}},
  \bibinfo {author} {\bibfnamefont {R.~F.~G.}\ \bibnamefont {Ruiz}}, \bibinfo {author} {\bibfnamefont {F.~P.}\ \bibnamefont {Gustafsson}}, \bibinfo {author} {\bibfnamefont {G.}~\bibnamefont {Hagen}}, \bibinfo {author} {\bibfnamefont {G.~R.}\ \bibnamefont {Jansen}}, \bibinfo {author} {\bibfnamefont {A.}~\bibnamefont {Kanellakopoulos}}, \bibinfo {author} {\bibfnamefont {M.}~\bibnamefont {Kortelainen}}, \bibinfo {author} {\bibfnamefont {W.}~\bibnamefont {Nazarewicz}}, \bibinfo {author} {\bibfnamefont {G.}~\bibnamefont {Neyens}}, \bibinfo {author} {\bibfnamefont {T.}~\bibnamefont {Papenbrock}}, \bibinfo {author} {\bibfnamefont {P.~G.}\ \bibnamefont {Reinhard}}, \bibinfo {author} {\bibfnamefont {C.~M.}\ \bibnamefont {Ricketts}}, \bibinfo {author} {\bibfnamefont {B.~K.}\ \bibnamefont {Sahoo}}, \bibinfo {author} {\bibfnamefont {A.~R.}\ \bibnamefont {Vernon}},\ and\ \bibinfo {author} {\bibfnamefont {S.~G.}\ \bibnamefont {Wilkins}},\ }\bibfield  {title} {\bibinfo {title} {Charge radii of exotic potassium isotopes
  challenge nuclear theory and the magic character of n = 32},\ }\href {https://doi.org/10.1038/s41567-020-01136-5} {\bibfield  {journal} {\bibinfo  {journal} {Nat. Phys.}\ }\textbf {\bibinfo {volume} {17}},\ \bibinfo {pages} {439} (\bibinfo {year} {2021})}\BibitemShut {NoStop}%
\bibitem [{\citenamefont {Blundell}\ \emph {et~al.}(1987)\citenamefont {Blundell}, \citenamefont {Baird}, \citenamefont {Palmer}, \citenamefont {Stacey},\ and\ \citenamefont {Woodgate}}]{S_A_Blundell_1987}%
  \BibitemOpen
  \bibfield  {author} {\bibinfo {author} {\bibfnamefont {S.~A.}\ \bibnamefont {Blundell}}, \bibinfo {author} {\bibfnamefont {P.~E.~G.}\ \bibnamefont {Baird}}, \bibinfo {author} {\bibfnamefont {C.~W.~P.}\ \bibnamefont {Palmer}}, \bibinfo {author} {\bibfnamefont {D.~N.}\ \bibnamefont {Stacey}},\ and\ \bibinfo {author} {\bibfnamefont {G.~K.}\ \bibnamefont {Woodgate}},\ }\href {https://doi.org/10.1088/0022-3700/20/15/015} {\bibfield  {journal} {\bibinfo  {journal} {J. Phys. B}\ }\textbf {\bibinfo {volume} {20}},\ \bibinfo {pages} {3663} (\bibinfo {year} {1987})}\BibitemShut {NoStop}%
\bibitem [{\citenamefont {Papoulia}\ \emph {et~al.}(2016)\citenamefont {Papoulia}, \citenamefont {Carlsson},\ and\ \citenamefont {Ekman}}]{PhysRevA.94.042502}%
  \BibitemOpen
  \bibfield  {author} {\bibinfo {author} {\bibfnamefont {A.}~\bibnamefont {Papoulia}}, \bibinfo {author} {\bibfnamefont {B.~G.}\ \bibnamefont {Carlsson}},\ and\ \bibinfo {author} {\bibfnamefont {J.}~\bibnamefont {Ekman}},\ }\href {https://doi.org/10.1103/PhysRevA.94.042502} {\bibfield  {journal} {\bibinfo  {journal} {Phys. Rev. A}\ }\textbf {\bibinfo {volume} {94}},\ \bibinfo {pages} {042502} (\bibinfo {year} {2016})}\BibitemShut {NoStop}%
\bibitem [{\citenamefont {Xie}\ \emph {et~al.}(2023{\natexlab{a}})\citenamefont {Xie}, \citenamefont {Naito}, \citenamefont {Li},\ and\ \citenamefont {Liang}}]{xie2023revisiting}%
  \BibitemOpen
  \bibfield  {author} {\bibinfo {author} {\bibfnamefont {H.~H.}\ \bibnamefont {Xie}}, \bibinfo {author} {\bibfnamefont {T.}~\bibnamefont {Naito}}, \bibinfo {author} {\bibfnamefont {J.}~\bibnamefont {Li}},\ and\ \bibinfo {author} {\bibfnamefont {H.}~\bibnamefont {Liang}},\ }\href@noop {} {} (\bibinfo {year} {2023}{\natexlab{a}}),\ \Eprint {https://arxiv.org/abs/2306.09026} {arXiv:2306.09026 [nucl-th]} \BibitemShut {NoStop}%
\bibitem [{\citenamefont {Ring}(1996)}]{RING1996193}%
  \BibitemOpen
  \bibfield  {author} {\bibinfo {author} {\bibfnamefont {P.}~\bibnamefont {Ring}},\ }\href {https://doi.org/https://doi.org/10.1016/0146-6410(96)00054-3} {\bibfield  {journal} {\bibinfo  {journal} {Prog. Part. Nucl. Phys.}\ }\textbf {\bibinfo {volume} {37}},\ \bibinfo {pages} {193} (\bibinfo {year} {1996})}\BibitemShut {NoStop}%
\bibitem [{\citenamefont {Meng}\ \emph {et~al.}(2006)\citenamefont {Meng}, \citenamefont {Toki}, \citenamefont {Zhou}, \citenamefont {Zhang}, \citenamefont {Long},\ and\ \citenamefont {Geng}}]{MENG2006470}%
  \BibitemOpen
  \bibfield  {author} {\bibinfo {author} {\bibfnamefont {J.}~\bibnamefont {Meng}}, \bibinfo {author} {\bibfnamefont {H.}~\bibnamefont {Toki}}, \bibinfo {author} {\bibfnamefont {S.~G.}\ \bibnamefont {Zhou}}, \bibinfo {author} {\bibfnamefont {S.~Q.}\ \bibnamefont {Zhang}}, \bibinfo {author} {\bibfnamefont {W.~H.}\ \bibnamefont {Long}},\ and\ \bibinfo {author} {\bibfnamefont {L.~S.}\ \bibnamefont {Geng}},\ }\href {https://doi.org/https://doi.org/10.1016/j.ppnp.2005.06.001} {\bibfield  {journal} {\bibinfo  {journal} {Prog. Part. Nucl. Phys.}\ }\textbf {\bibinfo {volume} {57}},\ \bibinfo {pages} {470} (\bibinfo {year} {2006})}\BibitemShut {NoStop}%
\bibitem [{\citenamefont {Meng}\ and\ \citenamefont {Zhou}(2015)}]{Meng_2015}%
  \BibitemOpen
  \bibfield  {author} {\bibinfo {author} {\bibfnamefont {J.}~\bibnamefont {Meng}}\ and\ \bibinfo {author} {\bibfnamefont {S.~G.}\ \bibnamefont {Zhou}},\ }\href {https://doi.org/10.1088/0954-3899/42/9/093101} {\bibfield  {journal} {\bibinfo  {journal} {J. Phys. G}\ }\textbf {\bibinfo {volume} {42}},\ \bibinfo {pages} {093101} (\bibinfo {year} {2015})}\BibitemShut {NoStop}%
\bibitem [{\citenamefont {Vretenar}\ \emph {et~al.}(2005)\citenamefont {Vretenar}, \citenamefont {Afanasjev}, \citenamefont {Lalazissis},\ and\ \citenamefont {Ring}}]{VRETENAR2005101}%
  \BibitemOpen
  \bibfield  {author} {\bibinfo {author} {\bibfnamefont {D.}~\bibnamefont {Vretenar}}, \bibinfo {author} {\bibfnamefont {A.}~\bibnamefont {Afanasjev}}, \bibinfo {author} {\bibfnamefont {G.}~\bibnamefont {Lalazissis}},\ and\ \bibinfo {author} {\bibfnamefont {P.}~\bibnamefont {Ring}},\ }\href {https://doi.org/https://doi.org/10.1016/j.physrep.2004.10.001} {\bibfield  {journal} {\bibinfo  {journal} {Phys. Rep.}\ }\textbf {\bibinfo {volume} {409}},\ \bibinfo {pages} {101} (\bibinfo {year} {2005})}\BibitemShut {NoStop}%
\bibitem [{\citenamefont {Nikšić}\ \emph {et~al.}(2011)\citenamefont {Nikšić}, \citenamefont {Vretenar},\ and\ \citenamefont {Ring}}]{NIKSIC2011519}%
  \BibitemOpen
  \bibfield  {author} {\bibinfo {author} {\bibfnamefont {T.}~\bibnamefont {Nikšić}}, \bibinfo {author} {\bibfnamefont {D.}~\bibnamefont {Vretenar}},\ and\ \bibinfo {author} {\bibfnamefont {P.}~\bibnamefont {Ring}},\ }\href {https://doi.org/https://doi.org/10.1016/j.ppnp.2011.01.055} {\bibfield  {journal} {\bibinfo  {journal} {Prog. Part. Nucl. Phys.}\ }\textbf {\bibinfo {volume} {66}},\ \bibinfo {pages} {519} (\bibinfo {year} {2011})}\BibitemShut {NoStop}%
\bibitem [{\citenamefont {Meng}\ \emph {et~al.}(2013)\citenamefont {Meng}, \citenamefont {Peng}, \citenamefont {Zhang},\ and\ \citenamefont {Zhao}}]{meng2013progress}%
  \BibitemOpen
  \bibfield  {author} {\bibinfo {author} {\bibfnamefont {J.}~\bibnamefont {Meng}}, \bibinfo {author} {\bibfnamefont {J.}~\bibnamefont {Peng}}, \bibinfo {author} {\bibfnamefont {S.-Q.}\ \bibnamefont {Zhang}},\ and\ \bibinfo {author} {\bibfnamefont {P.-W.}\ \bibnamefont {Zhao}},\ }\href {https://doi.org/10.1007/s11467-013-0287-y} {\bibfield  {journal} {\bibinfo  {journal} {Front. Phys.}\ }\textbf {\bibinfo {volume} {8}},\ \bibinfo {pages} {55} (\bibinfo {year} {2013})}\BibitemShut {NoStop}%
\bibitem [{\citenamefont {Sharma}\ \emph {et~al.}(1993)\citenamefont {Sharma}, \citenamefont {Lalazissis},\ and\ \citenamefont {Ring}}]{SHARMA19939}%
  \BibitemOpen
  \bibfield  {author} {\bibinfo {author} {\bibfnamefont {M.~M.}\ \bibnamefont {Sharma}}, \bibinfo {author} {\bibfnamefont {G.~A.}\ \bibnamefont {Lalazissis}},\ and\ \bibinfo {author} {\bibfnamefont {P.}~\bibnamefont {Ring}},\ }\href {https://doi.org/https://doi.org/10.1016/0370-2693(93)91561-Z} {\bibfield  {journal} {\bibinfo  {journal} {Phys. Lett. B}\ }\textbf {\bibinfo {volume} {317}},\ \bibinfo {pages} {9} (\bibinfo {year} {1993})}\BibitemShut {NoStop}%
\bibitem [{\citenamefont {Naito}\ \emph {et~al.}(2023)\citenamefont {Naito}, \citenamefont {Oishi}, \citenamefont {Sagawa},\ and\ \citenamefont {Wang}}]{PhysRevC.107.054307}%
  \BibitemOpen
  \bibfield  {author} {\bibinfo {author} {\bibfnamefont {T.}~\bibnamefont {Naito}}, \bibinfo {author} {\bibfnamefont {T.}~\bibnamefont {Oishi}}, \bibinfo {author} {\bibfnamefont {H.}~\bibnamefont {Sagawa}},\ and\ \bibinfo {author} {\bibfnamefont {Z.}~\bibnamefont {Wang}},\ }\href {https://doi.org/10.1103/PhysRevC.107.054307} {\bibfield  {journal} {\bibinfo  {journal} {Phys. Rev. C}\ }\textbf {\bibinfo {volume} {107}},\ \bibinfo {pages} {054307} (\bibinfo {year} {2023})}\BibitemShut {NoStop}%
\bibitem [{\citenamefont {Liang}\ \emph {et~al.}(2015)\citenamefont {Liang}, \citenamefont {Meng},\ and\ \citenamefont {Zhou}}]{LIANG20151}%
  \BibitemOpen
  \bibfield  {author} {\bibinfo {author} {\bibfnamefont {H.}~\bibnamefont {Liang}}, \bibinfo {author} {\bibfnamefont {J.}~\bibnamefont {Meng}},\ and\ \bibinfo {author} {\bibfnamefont {S.-G.}\ \bibnamefont {Zhou}},\ }\href {https://doi.org/https://doi.org/10.1016/j.physrep.2014.12.005} {\bibfield  {journal} {\bibinfo  {journal} {Phys. Rep.}\ }\textbf {\bibinfo {volume} {570}},\ \bibinfo {pages} {1} (\bibinfo {year} {2015})}\BibitemShut {NoStop}%
\bibitem [{\citenamefont {Zhou}\ \emph {et~al.}(2003)\citenamefont {Zhou}, \citenamefont {Meng},\ and\ \citenamefont {Ring}}]{PhysRevLett.91.262501}%
  \BibitemOpen
  \bibfield  {author} {\bibinfo {author} {\bibfnamefont {S.-G.}\ \bibnamefont {Zhou}}, \bibinfo {author} {\bibfnamefont {J.}~\bibnamefont {Meng}},\ and\ \bibinfo {author} {\bibfnamefont {P.}~\bibnamefont {Ring}},\ }\href {https://doi.org/10.1103/PhysRevLett.91.262501} {\bibfield  {journal} {\bibinfo  {journal} {Phys. Rev. Lett.}\ }\textbf {\bibinfo {volume} {91}},\ \bibinfo {pages} {262501} (\bibinfo {year} {2003})}\BibitemShut {NoStop}%
\bibitem [{\citenamefont {Liang}\ \emph {et~al.}(2010)\citenamefont {Liang}, \citenamefont {Long}, \citenamefont {Meng},\ and\ \citenamefont {Van~Giai}}]{liang2010spin}%
  \BibitemOpen
  \bibfield  {author} {\bibinfo {author} {\bibfnamefont {H.}~\bibnamefont {Liang}}, \bibinfo {author} {\bibfnamefont {W.~H.}\ \bibnamefont {Long}}, \bibinfo {author} {\bibfnamefont {J.}~\bibnamefont {Meng}},\ and\ \bibinfo {author} {\bibfnamefont {N.}~\bibnamefont {Van~Giai}},\ }\href {https://doi.org/10.1140/epja/i2010-10938-6} {\bibfield  {journal} {\bibinfo  {journal} {Eur. Phys. J. A}\ }\textbf {\bibinfo {volume} {44}},\ \bibinfo {pages} {119} (\bibinfo {year} {2010})}\BibitemShut {NoStop}%
\bibitem [{\citenamefont {Li}\ and\ \citenamefont {Sun}(2020)}]{Li_2020}%
  \BibitemOpen
  \bibfield  {author} {\bibinfo {author} {\bibfnamefont {J.}~\bibnamefont {Li}}\ and\ \bibinfo {author} {\bibfnamefont {W.-J.}\ \bibnamefont {Sun}},\ }\href {https://doi.org/10.1088/1572-9494/ab7708} {\bibfield  {journal} {\bibinfo  {journal} {Commun. in Theor. Phys.}\ }\textbf {\bibinfo {volume} {72}},\ \bibinfo {pages} {055301} (\bibinfo {year} {2020})}\BibitemShut {NoStop}%
\bibitem [{\citenamefont {Li}\ \emph {et~al.}(2011{\natexlab{a}})\citenamefont {Li}, \citenamefont {Meng}, \citenamefont {Ring}, \citenamefont {Yao},\ and\ \citenamefont {Arima}}]{Li2011}%
  \BibitemOpen
  \bibfield  {author} {\bibinfo {author} {\bibfnamefont {J.}~\bibnamefont {Li}}, \bibinfo {author} {\bibfnamefont {J.}~\bibnamefont {Meng}}, \bibinfo {author} {\bibfnamefont {P.}~\bibnamefont {Ring}}, \bibinfo {author} {\bibfnamefont {J.}~\bibnamefont {Yao}},\ and\ \bibinfo {author} {\bibfnamefont {A.}~\bibnamefont {Arima}},\ }\href {https://doi.org/10.1007/s11433-010-4215-7} {\bibfield  {journal} {\bibinfo  {journal} {Sci. China Phys. Mech. Astron.}\ }\textbf {\bibinfo {volume} {54}},\ \bibinfo {pages} {204} (\bibinfo {year} {2011}{\natexlab{a}})}\BibitemShut {NoStop}%
\bibitem [{\citenamefont {Li}\ \emph {et~al.}(2011{\natexlab{b}})\citenamefont {Li}, \citenamefont {Yao}, \citenamefont {Meng},\ and\ \citenamefont {Arima}}]{10.1143/PTP.125.1185}%
  \BibitemOpen
  \bibfield  {author} {\bibinfo {author} {\bibfnamefont {J.}~\bibnamefont {Li}}, \bibinfo {author} {\bibfnamefont {J.~M.}\ \bibnamefont {Yao}}, \bibinfo {author} {\bibfnamefont {J.}~\bibnamefont {Meng}},\ and\ \bibinfo {author} {\bibfnamefont {A.}~\bibnamefont {Arima}},\ }\href {https://doi.org/10.1143/PTP.125.1185} {\bibfield  {journal} {\bibinfo  {journal} {Prog. Theor. Phys.}\ }\textbf {\bibinfo {volume} {125}},\ \bibinfo {pages} {1185} (\bibinfo {year} {2011}{\natexlab{b}})}\BibitemShut {NoStop}%
\bibitem [{\citenamefont {Wei}\ \emph {et~al.}(2012)\citenamefont {Wei}, \citenamefont {Li},\ and\ \citenamefont {Meng}}]{10.1143/PTPS.196.400}%
  \BibitemOpen
  \bibfield  {author} {\bibinfo {author} {\bibfnamefont {J.}~\bibnamefont {Wei}}, \bibinfo {author} {\bibfnamefont {J.}~\bibnamefont {Li}},\ and\ \bibinfo {author} {\bibfnamefont {J.}~\bibnamefont {Meng}},\ }\href {https://doi.org/10.1143/PTPS.196.400} {\bibfield  {journal} {\bibinfo  {journal} {Prog. Theor. Phys. Suppl.}\ }\textbf {\bibinfo {volume} {196}},\ \bibinfo {pages} {400} (\bibinfo {year} {2012})}\BibitemShut {NoStop}%
\bibitem [{\citenamefont {Li}\ \emph {et~al.}(2013{\natexlab{a}})\citenamefont {Li}, \citenamefont {Wei}, \citenamefont {Hu}, \citenamefont {Ring},\ and\ \citenamefont {Meng}}]{PhysRevC.88.064307}%
  \BibitemOpen
  \bibfield  {author} {\bibinfo {author} {\bibfnamefont {J.}~\bibnamefont {Li}}, \bibinfo {author} {\bibfnamefont {J.~X.}\ \bibnamefont {Wei}}, \bibinfo {author} {\bibfnamefont {J.~N.}\ \bibnamefont {Hu}}, \bibinfo {author} {\bibfnamefont {P.}~\bibnamefont {Ring}},\ and\ \bibinfo {author} {\bibfnamefont {J.}~\bibnamefont {Meng}},\ }\href {https://doi.org/10.1103/PhysRevC.88.064307} {\bibfield  {journal} {\bibinfo  {journal} {Phys. Rev. C}\ }\textbf {\bibinfo {volume} {88}},\ \bibinfo {pages} {064307} (\bibinfo {year} {2013}{\natexlab{a}})}\BibitemShut {NoStop}%
\bibitem [{\citenamefont {K\"onig}\ and\ \citenamefont {Ring}(1993)}]{PhysRevLett.71.3079}%
  \BibitemOpen
  \bibfield  {author} {\bibinfo {author} {\bibfnamefont {J.}~\bibnamefont {K\"onig}}\ and\ \bibinfo {author} {\bibfnamefont {P.}~\bibnamefont {Ring}},\ }\href {https://doi.org/10.1103/PhysRevLett.71.3079} {\bibfield  {journal} {\bibinfo  {journal} {Phys. Rev. Lett.}\ }\textbf {\bibinfo {volume} {71}},\ \bibinfo {pages} {3079} (\bibinfo {year} {1993})}\BibitemShut {NoStop}%
\bibitem [{\citenamefont {Zhao}\ \emph {et~al.}(2011)\citenamefont {Zhao}, \citenamefont {Peng}, \citenamefont {Liang}, \citenamefont {Ring},\ and\ \citenamefont {Meng}}]{PhysRevLett.107.122501}%
  \BibitemOpen
  \bibfield  {author} {\bibinfo {author} {\bibfnamefont {P.~W.}\ \bibnamefont {Zhao}}, \bibinfo {author} {\bibfnamefont {J.}~\bibnamefont {Peng}}, \bibinfo {author} {\bibfnamefont {H.~Z.}\ \bibnamefont {Liang}}, \bibinfo {author} {\bibfnamefont {P.}~\bibnamefont {Ring}},\ and\ \bibinfo {author} {\bibfnamefont {J.}~\bibnamefont {Meng}},\ }\href {https://doi.org/10.1103/PhysRevLett.107.122501} {\bibfield  {journal} {\bibinfo  {journal} {Phys. Rev. Lett.}\ }\textbf {\bibinfo {volume} {107}},\ \bibinfo {pages} {122501} (\bibinfo {year} {2011})}\BibitemShut {NoStop}%
\bibitem [{\citenamefont {Meng}\ and\ \citenamefont {Ring}(1996)}]{PhysRevLett.77.3963}%
  \BibitemOpen
  \bibfield  {author} {\bibinfo {author} {\bibfnamefont {J.}~\bibnamefont {Meng}}\ and\ \bibinfo {author} {\bibfnamefont {P.}~\bibnamefont {Ring}},\ }\href {https://doi.org/10.1103/PhysRevLett.77.3963} {\bibfield  {journal} {\bibinfo  {journal} {Phys. Rev. Lett.}\ }\textbf {\bibinfo {volume} {77}},\ \bibinfo {pages} {3963} (\bibinfo {year} {1996})}\BibitemShut {NoStop}%
\bibitem [{\citenamefont {Meng}(1998)}]{Meng1998NPA}%
  \BibitemOpen
  \bibfield  {author} {\bibinfo {author} {\bibfnamefont {J.}~\bibnamefont {Meng}},\ }\href {https://doi.org/https://doi.org/10.1016/S0375-9474(98)00178-X} {\bibfield  {journal} {\bibinfo  {journal} {Nucl. Phys. A}\ }\textbf {\bibinfo {volume} {635}},\ \bibinfo {pages} {3} (\bibinfo {year} {1998})}\BibitemShut {NoStop}%
\bibitem [{\citenamefont {Meng}\ and\ \citenamefont {Ring}(1998)}]{PhysRevLett.80.460}%
  \BibitemOpen
  \bibfield  {author} {\bibinfo {author} {\bibfnamefont {J.}~\bibnamefont {Meng}}\ and\ \bibinfo {author} {\bibfnamefont {P.}~\bibnamefont {Ring}},\ }\href {https://doi.org/10.1103/PhysRevLett.80.460} {\bibfield  {journal} {\bibinfo  {journal} {Phys. Rev. Lett.}\ }\textbf {\bibinfo {volume} {80}},\ \bibinfo {pages} {460} (\bibinfo {year} {1998})}\BibitemShut {NoStop}%
\bibitem [{\citenamefont {Meng}\ \emph {et~al.}(2002{\natexlab{a}})\citenamefont {Meng}, \citenamefont {Toki}, \citenamefont {Zeng}, \citenamefont {Zhang},\ and\ \citenamefont {Zhou}}]{PhysRevC.65.041302}%
  \BibitemOpen
  \bibfield  {author} {\bibinfo {author} {\bibfnamefont {J.}~\bibnamefont {Meng}}, \bibinfo {author} {\bibfnamefont {H.}~\bibnamefont {Toki}}, \bibinfo {author} {\bibfnamefont {J.~Y.}\ \bibnamefont {Zeng}}, \bibinfo {author} {\bibfnamefont {S.~Q.}\ \bibnamefont {Zhang}},\ and\ \bibinfo {author} {\bibfnamefont {S.-G.}\ \bibnamefont {Zhou}},\ }\href {https://doi.org/10.1103/PhysRevC.65.041302} {\bibfield  {journal} {\bibinfo  {journal} {Phys. Rev. C}\ }\textbf {\bibinfo {volume} {65}},\ \bibinfo {pages} {041302} (\bibinfo {year} {2002}{\natexlab{a}})}\BibitemShut {NoStop}%
\bibitem [{\citenamefont {Zhang}\ \emph {et~al.}(2002)\citenamefont {Zhang}, \citenamefont {Meng}, \citenamefont {Zhou},\ and\ \citenamefont {Zeng}}]{Shuang_Quan_2002}%
  \BibitemOpen
  \bibfield  {author} {\bibinfo {author} {\bibfnamefont {S.-Q.}\ \bibnamefont {Zhang}}, \bibinfo {author} {\bibfnamefont {J.}~\bibnamefont {Meng}}, \bibinfo {author} {\bibfnamefont {S.-G.}\ \bibnamefont {Zhou}},\ and\ \bibinfo {author} {\bibfnamefont {J.-Y.}\ \bibnamefont {Zeng}},\ }\href {https://doi.org/10.1088/0256-307x/19/3/308} {\bibfield  {journal} {\bibinfo  {journal} {Chinese Phys. Lett.}\ }\textbf {\bibinfo {volume} {19}},\ \bibinfo {pages} {312} (\bibinfo {year} {2002})}\BibitemShut {NoStop}%
\bibitem [{\citenamefont {Meng}\ \emph {et~al.}(1998)\citenamefont {Meng}, \citenamefont {Tanihata},\ and\ \citenamefont {Yamaji}}]{MENG19981}%
  \BibitemOpen
  \bibfield  {author} {\bibinfo {author} {\bibfnamefont {J.}~\bibnamefont {Meng}}, \bibinfo {author} {\bibfnamefont {I.}~\bibnamefont {Tanihata}},\ and\ \bibinfo {author} {\bibfnamefont {S.}~\bibnamefont {Yamaji}},\ }\href {https://doi.org/https://doi.org/10.1016/S0370-2693(97)01386-5} {\bibfield  {journal} {\bibinfo  {journal} {Phys. Lett. B}\ }\textbf {\bibinfo {volume} {419}},\ \bibinfo {pages} {1} (\bibinfo {year} {1998})}\BibitemShut {NoStop}%
\bibitem [{\citenamefont {Meng}\ \emph {et~al.}(2002{\natexlab{b}})\citenamefont {Meng}, \citenamefont {Zhou},\ and\ \citenamefont {Tanihata}}]{MENG2002209}%
  \BibitemOpen
  \bibfield  {author} {\bibinfo {author} {\bibfnamefont {J.}~\bibnamefont {Meng}}, \bibinfo {author} {\bibfnamefont {S.-G.}\ \bibnamefont {Zhou}},\ and\ \bibinfo {author} {\bibfnamefont {I.}~\bibnamefont {Tanihata}},\ }\href {https://doi.org/https://doi.org/10.1016/S0370-2693(02)01574-5} {\bibfield  {journal} {\bibinfo  {journal} {Phys. Lett. B}\ }\textbf {\bibinfo {volume} {532}},\ \bibinfo {pages} {209} (\bibinfo {year} {2002}{\natexlab{b}})}\BibitemShut {NoStop}%
\bibitem [{\citenamefont {Zhao}\ \emph {et~al.}(2010)\citenamefont {Zhao}, \citenamefont {Li}, \citenamefont {Yao},\ and\ \citenamefont {Meng}}]{PhysRevC.82.054319}%
  \BibitemOpen
  \bibfield  {author} {\bibinfo {author} {\bibfnamefont {P.~W.}\ \bibnamefont {Zhao}}, \bibinfo {author} {\bibfnamefont {Z.~P.}\ \bibnamefont {Li}}, \bibinfo {author} {\bibfnamefont {J.~M.}\ \bibnamefont {Yao}},\ and\ \bibinfo {author} {\bibfnamefont {J.}~\bibnamefont {Meng}},\ }\href {https://doi.org/10.1103/PhysRevC.82.054319} {\bibfield  {journal} {\bibinfo  {journal} {Phys. Rev. C}\ }\textbf {\bibinfo {volume} {82}},\ \bibinfo {pages} {054319} (\bibinfo {year} {2010})}\BibitemShut {NoStop}%
\bibitem [{\citenamefont {Xia}\ \emph {et~al.}(2018)\citenamefont {Xia}, \citenamefont {Lim}, \citenamefont {Zhao}, \citenamefont {Liang}, \citenamefont {Qu}, \citenamefont {Chen}, \citenamefont {Liu}, \citenamefont {Zhang}, \citenamefont {Zhang}, \citenamefont {Kim},\ and\ \citenamefont {Meng}}]{Xia2018}%
  \BibitemOpen
  \bibfield  {author} {\bibinfo {author} {\bibfnamefont {X.~W.}\ \bibnamefont {Xia}}, \bibinfo {author} {\bibfnamefont {Y.}~\bibnamefont {Lim}}, \bibinfo {author} {\bibfnamefont {P.~W.}\ \bibnamefont {Zhao}}, \bibinfo {author} {\bibfnamefont {H.~Z.}\ \bibnamefont {Liang}}, \bibinfo {author} {\bibfnamefont {X.~Y.}\ \bibnamefont {Qu}}, \bibinfo {author} {\bibfnamefont {Y.}~\bibnamefont {Chen}}, \bibinfo {author} {\bibfnamefont {H.}~\bibnamefont {Liu}}, \bibinfo {author} {\bibfnamefont {L.~F.}\ \bibnamefont {Zhang}}, \bibinfo {author} {\bibfnamefont {S.~Q.}\ \bibnamefont {Zhang}}, \bibinfo {author} {\bibfnamefont {Y.}~\bibnamefont {Kim}},\ and\ \bibinfo {author} {\bibfnamefont {J.}~\bibnamefont {Meng}},\ }\href {https://doi.org/10.1016/j.adt.2017.09.001} {\bibfield  {journal} {\bibinfo  {journal} {At. Data Nucl. Data Tables}\ }\textbf {\bibinfo {volume} {121-122}},\ \bibinfo {pages} {1} (\bibinfo {year} {2018})}\BibitemShut {NoStop}%
\bibitem [{\citenamefont {Meng}(2015)}]{Meng2015}%
  \BibitemOpen
  \bibfield  {author} {\bibinfo {author} {\bibfnamefont {J.}~\bibnamefont {Meng}},\ }\href {https://doi.org/10.1142/9872} {\emph {\bibinfo {title} {Relativistic Density Functional for Nuclear Structure}}}\ (\bibinfo  {publisher} {World Scientific, Singapore},\ \bibinfo {year} {2015})\BibitemShut {NoStop}%
\bibitem [{\citenamefont {Friar}\ and\ \citenamefont {Negele}(1975)}]{Friar1975}%
  \BibitemOpen
  \bibfield  {author} {\bibinfo {author} {\bibfnamefont {J.~L.}\ \bibnamefont {Friar}}\ and\ \bibinfo {author} {\bibfnamefont {J.~W.}\ \bibnamefont {Negele}},\ }\href {https://doi.org/10.1007/978-1-4757-4398-2_3} {\bibfield  {journal} {\bibinfo  {journal} {Adv. Nucl. Phys.}\ }\textbf {\bibinfo {volume} {8}},\ \bibinfo {pages} {219} (\bibinfo {year} {1975})}\BibitemShut {NoStop}%
\bibitem [{\citenamefont {Kurasawa}\ and\ \citenamefont {Suzuki}(2000)}]{PhysRevC.62.054303}%
  \BibitemOpen
  \bibfield  {author} {\bibinfo {author} {\bibfnamefont {H.}~\bibnamefont {Kurasawa}}\ and\ \bibinfo {author} {\bibfnamefont {T.}~\bibnamefont {Suzuki}},\ }\href {https://doi.org/10.1103/PhysRevC.62.054303} {\bibfield  {journal} {\bibinfo  {journal} {Phys. Rev. C}\ }\textbf {\bibinfo {volume} {62}},\ \bibinfo {pages} {054303} (\bibinfo {year} {2000})}\BibitemShut {NoStop}%
\bibitem [{\citenamefont {Reinhard}\ and\ \citenamefont {Nazarewicz}(2021)}]{PhysRevC.103.054310}%
  \BibitemOpen
  \bibfield  {author} {\bibinfo {author} {\bibfnamefont {P.-G.}\ \bibnamefont {Reinhard}}\ and\ \bibinfo {author} {\bibfnamefont {W.}~\bibnamefont {Nazarewicz}},\ }\href {https://doi.org/10.1103/PhysRevC.103.054310} {\bibfield  {journal} {\bibinfo  {journal} {Phys. Rev. C}\ }\textbf {\bibinfo {volume} {103}},\ \bibinfo {pages} {054310} (\bibinfo {year} {2021})}\BibitemShut {NoStop}%
\bibitem [{\citenamefont {Kurasawa}\ and\ \citenamefont {Suzuki}(2019)}]{kurasawa2019n}%
  \BibitemOpen
  \bibfield  {author} {\bibinfo {author} {\bibfnamefont {H.}~\bibnamefont {Kurasawa}}\ and\ \bibinfo {author} {\bibfnamefont {T.}~\bibnamefont {Suzuki}},\ }\href {https://doi.org/10.1093/ptep/ptz121} {\bibfield  {journal} {\bibinfo  {journal} {Prog. Theor. Exp. Phys.}\ }\textbf {\bibinfo {volume} {2019}},\ \bibinfo {pages} {113D01} (\bibinfo {year} {2019})}\BibitemShut {NoStop}%
\bibitem [{\citenamefont {Tiesinga}\ \emph {et~al.}(2021)\citenamefont {Tiesinga}, \citenamefont {Mohr}, \citenamefont {Newell},\ and\ \citenamefont {Taylor}}]{RevModPhys.93.025010}%
  \BibitemOpen
  \bibfield  {author} {\bibinfo {author} {\bibfnamefont {E.}~\bibnamefont {Tiesinga}}, \bibinfo {author} {\bibfnamefont {P.~J.}\ \bibnamefont {Mohr}}, \bibinfo {author} {\bibfnamefont {D.~B.}\ \bibnamefont {Newell}},\ and\ \bibinfo {author} {\bibfnamefont {B.~N.}\ \bibnamefont {Taylor}},\ }\href {https://doi.org/10.1103/RevModPhys.93.025010} {\bibfield  {journal} {\bibinfo  {journal} {Rev. Mod. Phys.}\ }\textbf {\bibinfo {volume} {93}},\ \bibinfo {pages} {025010} (\bibinfo {year} {2021})}\BibitemShut {NoStop}%
\bibitem [{\citenamefont {Atac}\ \emph {et~al.}(2021)\citenamefont {Atac}, \citenamefont {Constantinou}, \citenamefont {Meziani}, \citenamefont {Paolone},\ and\ \citenamefont {Sparveris}}]{atac2021measurement}%
  \BibitemOpen
  \bibfield  {author} {\bibinfo {author} {\bibfnamefont {H.}~\bibnamefont {Atac}}, \bibinfo {author} {\bibfnamefont {M.}~\bibnamefont {Constantinou}}, \bibinfo {author} {\bibfnamefont {Z.-E.}\ \bibnamefont {Meziani}}, \bibinfo {author} {\bibfnamefont {M.}~\bibnamefont {Paolone}},\ and\ \bibinfo {author} {\bibfnamefont {N.}~\bibnamefont {Sparveris}},\ }\href {https://doi.org/10.1038/s41467-021-22028-z} {\bibfield  {journal} {\bibinfo  {journal} {Nat. Commun.}\ }\textbf {\bibinfo {volume} {12}},\ \bibinfo {pages} {1} (\bibinfo {year} {2021})}\BibitemShut {NoStop}%
\bibitem [{\citenamefont {Xie}\ \emph {et~al.}(2023{\natexlab{b}})\citenamefont {Xie}, \citenamefont {Li}, \citenamefont {Jiao},\ and\ \citenamefont {Ho}}]{PhysRevA.107.042807}%
  \BibitemOpen
  \bibfield  {author} {\bibinfo {author} {\bibfnamefont {H.~H.}\ \bibnamefont {Xie}}, \bibinfo {author} {\bibfnamefont {J.}~\bibnamefont {Li}}, \bibinfo {author} {\bibfnamefont {L.~G.}\ \bibnamefont {Jiao}},\ and\ \bibinfo {author} {\bibfnamefont {Y.~K.}\ \bibnamefont {Ho}},\ }\href {https://doi.org/10.1103/PhysRevA.107.042807} {\bibfield  {journal} {\bibinfo  {journal} {Phys. Rev. A}\ }\textbf {\bibinfo {volume} {107}},\ \bibinfo {pages} {042807} (\bibinfo {year} {2023}{\natexlab{b}})}\BibitemShut {NoStop}%
\bibitem [{\citenamefont {Xie}\ and\ \citenamefont {Li}(2023)}]{xie2023impact}%
  \BibitemOpen
  \bibfield  {author} {\bibinfo {author} {\bibfnamefont {H.}~\bibnamefont {Xie}}\ and\ \bibinfo {author} {\bibfnamefont {J.}~\bibnamefont {Li}},\ }\href@noop {} {} (\bibinfo {year} {2023}),\ \Eprint {https://arxiv.org/abs/2308.02309} {arXiv:2308.02309 [nucl-th]} \BibitemShut {NoStop}%
\bibitem [{\citenamefont {Michel}\ \emph {et~al.}(2017)\citenamefont {Michel}, \citenamefont {Oreshkina},\ and\ \citenamefont {Keitel}}]{PhysRevA.96.032510}%
  \BibitemOpen
  \bibfield  {author} {\bibinfo {author} {\bibfnamefont {N.}~\bibnamefont {Michel}}, \bibinfo {author} {\bibfnamefont {N.~S.}\ \bibnamefont {Oreshkina}},\ and\ \bibinfo {author} {\bibfnamefont {C.~H.}\ \bibnamefont {Keitel}},\ }\href {https://doi.org/10.1103/PhysRevA.96.032510} {\bibfield  {journal} {\bibinfo  {journal} {Phys. Rev. A}\ }\textbf {\bibinfo {volume} {96}},\ \bibinfo {pages} {032510} (\bibinfo {year} {2017})}\BibitemShut {NoStop}%
\bibitem [{\citenamefont {Wang}\ \emph {et~al.}(2021)\citenamefont {Wang}, \citenamefont {Huang}, \citenamefont {Kondev}, \citenamefont {Audi},\ and\ \citenamefont {Naimi}}]{Wang_2021}%
  \BibitemOpen
  \bibfield  {author} {\bibinfo {author} {\bibfnamefont {M.}~\bibnamefont {Wang}}, \bibinfo {author} {\bibfnamefont {W.}~\bibnamefont {Huang}}, \bibinfo {author} {\bibfnamefont {F.}~\bibnamefont {Kondev}}, \bibinfo {author} {\bibfnamefont {G.}~\bibnamefont {Audi}},\ and\ \bibinfo {author} {\bibfnamefont {S.}~\bibnamefont {Naimi}},\ }\href {https://doi.org/10.1088/1674-1137/abddaf} {\bibfield  {journal} {\bibinfo  {journal} {Chinese Phys. C}\ }\textbf {\bibinfo {volume} {45}},\ \bibinfo {pages} {030003} (\bibinfo {year} {2021})}\BibitemShut {NoStop}%
\bibitem [{\citenamefont {Engel}(2002)}]{engel2002relativistic}%
  \BibitemOpen
  \bibfield  {author} {\bibinfo {author} {\bibfnamefont {E.}~\bibnamefont {Engel}},\ }\href@noop {} {\emph {\bibinfo {title} {Relativistic Electronic Structure Theory, Part 1. Fundamentals}}}\ (\bibinfo  {publisher} {Elsevier, Amsterdam},\ \bibinfo {year} {2002})\ pp.\ \bibinfo {pages} {524--624}\BibitemShut {NoStop}%
\bibitem [{\citenamefont {Zhang}\ \emph {et~al.}(2014)\citenamefont {Zhang}, \citenamefont {Niu}, \citenamefont {Li}, \citenamefont {Yao},\ and\ \citenamefont {Meng}}]{RN7}%
  \BibitemOpen
  \bibfield  {author} {\bibinfo {author} {\bibfnamefont {Q.-S.}\ \bibnamefont {Zhang}}, \bibinfo {author} {\bibfnamefont {Z.-M.}\ \bibnamefont {Niu}}, \bibinfo {author} {\bibfnamefont {Z.-P.}\ \bibnamefont {Li}}, \bibinfo {author} {\bibfnamefont {J.-M.}\ \bibnamefont {Yao}},\ and\ \bibinfo {author} {\bibfnamefont {J.}~\bibnamefont {Meng}},\ }\href {https://doi.org/10.1007/s11467-014-0413-5} {\bibfield  {journal} {\bibinfo  {journal} {Front. Phys.}\ }\textbf {\bibinfo {volume} {9}},\ \bibinfo {pages} {529} (\bibinfo {year} {2014})}\BibitemShut {NoStop}%
\bibitem [{\citenamefont {Lu}\ \emph {et~al.}(2015)\citenamefont {Lu}, \citenamefont {Li}, \citenamefont {Li}, \citenamefont {Yao},\ and\ \citenamefont {Meng}}]{PhysRevC.91.027304}%
  \BibitemOpen
  \bibfield  {author} {\bibinfo {author} {\bibfnamefont {K.~Q.}\ \bibnamefont {Lu}}, \bibinfo {author} {\bibfnamefont {Z.~X.}\ \bibnamefont {Li}}, \bibinfo {author} {\bibfnamefont {Z.~P.}\ \bibnamefont {Li}}, \bibinfo {author} {\bibfnamefont {J.~M.}\ \bibnamefont {Yao}},\ and\ \bibinfo {author} {\bibfnamefont {J.}~\bibnamefont {Meng}},\ }\href {https://doi.org/10.1103/PhysRevC.91.027304} {\bibfield  {journal} {\bibinfo  {journal} {Phys. Rev. C}\ }\textbf {\bibinfo {volume} {91}},\ \bibinfo {pages} {027304} (\bibinfo {year} {2015})}\BibitemShut {NoStop}%
\bibitem [{\citenamefont {Zhao}\ \emph {et~al.}(2012)\citenamefont {Zhao}, \citenamefont {Song}, \citenamefont {Sun}, \citenamefont {Geissel},\ and\ \citenamefont {Meng}}]{PhysRevC.86.064324}%
  \BibitemOpen
  \bibfield  {author} {\bibinfo {author} {\bibfnamefont {P.~W.}\ \bibnamefont {Zhao}}, \bibinfo {author} {\bibfnamefont {L.~S.}\ \bibnamefont {Song}}, \bibinfo {author} {\bibfnamefont {B.}~\bibnamefont {Sun}}, \bibinfo {author} {\bibfnamefont {H.}~\bibnamefont {Geissel}},\ and\ \bibinfo {author} {\bibfnamefont {J.}~\bibnamefont {Meng}},\ }\href {https://doi.org/10.1103/PhysRevC.86.064324} {\bibfield  {journal} {\bibinfo  {journal} {Phys. Rev. C}\ }\textbf {\bibinfo {volume} {86}},\ \bibinfo {pages} {064324} (\bibinfo {year} {2012})}\BibitemShut {NoStop}%
\bibitem [{\citenamefont {Yao}\ \emph {et~al.}(2013)\citenamefont {Yao}, \citenamefont {Mei},\ and\ \citenamefont {Li}}]{YAO2013459}%
  \BibitemOpen
  \bibfield  {author} {\bibinfo {author} {\bibfnamefont {J.}~\bibnamefont {Yao}}, \bibinfo {author} {\bibfnamefont {H.}~\bibnamefont {Mei}},\ and\ \bibinfo {author} {\bibfnamefont {Z.}~\bibnamefont {Li}},\ }\href {https://doi.org/https://doi.org/10.1016/j.physletb.2013.05.049} {\bibfield  {journal} {\bibinfo  {journal} {Phys. Lett. B}\ }\textbf {\bibinfo {volume} {723}},\ \bibinfo {pages} {459} (\bibinfo {year} {2013})}\BibitemShut {NoStop}%
\bibitem [{\citenamefont {Li}\ \emph {et~al.}(2013{\natexlab{b}})\citenamefont {Li}, \citenamefont {Song}, \citenamefont {Yao}, \citenamefont {Vretenar},\ and\ \citenamefont {Meng}}]{LI2013866}%
  \BibitemOpen
  \bibfield  {author} {\bibinfo {author} {\bibfnamefont {Z.}~\bibnamefont {Li}}, \bibinfo {author} {\bibfnamefont {B.}~\bibnamefont {Song}}, \bibinfo {author} {\bibfnamefont {J.}~\bibnamefont {Yao}}, \bibinfo {author} {\bibfnamefont {D.}~\bibnamefont {Vretenar}},\ and\ \bibinfo {author} {\bibfnamefont {J.}~\bibnamefont {Meng}},\ }\href {https://doi.org/https://doi.org/10.1016/j.physletb.2013.09.035} {\bibfield  {journal} {\bibinfo  {journal} {Phys. Lett. B}\ }\textbf {\bibinfo {volume} {726}},\ \bibinfo {pages} {866} (\bibinfo {year} {2013}{\natexlab{b}})}\BibitemShut {NoStop}%
\bibitem [{\citenamefont {Li}\ \emph {et~al.}(2012)\citenamefont {Li}, \citenamefont {Li}, \citenamefont {Xiang}, \citenamefont {Yao},\ and\ \citenamefont {Meng}}]{LI2012470}%
  \BibitemOpen
  \bibfield  {author} {\bibinfo {author} {\bibfnamefont {Z.}~\bibnamefont {Li}}, \bibinfo {author} {\bibfnamefont {C.}~\bibnamefont {Li}}, \bibinfo {author} {\bibfnamefont {J.}~\bibnamefont {Xiang}}, \bibinfo {author} {\bibfnamefont {J.}~\bibnamefont {Yao}},\ and\ \bibinfo {author} {\bibfnamefont {J.}~\bibnamefont {Meng}},\ }\href {https://doi.org/https://doi.org/10.1016/j.physletb.2012.09.061} {\bibfield  {journal} {\bibinfo  {journal} {Phys. Lett. B}\ }\textbf {\bibinfo {volume} {717}},\ \bibinfo {pages} {470} (\bibinfo {year} {2012})}\BibitemShut {NoStop}%
\bibitem [{\citenamefont {Xiang}\ \emph {et~al.}(2013)\citenamefont {Xiang}, \citenamefont {Li}, \citenamefont {Yao}, \citenamefont {Long}, \citenamefont {Ring},\ and\ \citenamefont {Meng}}]{PhysRevC.88.057301}%
  \BibitemOpen
  \bibfield  {author} {\bibinfo {author} {\bibfnamefont {J.}~\bibnamefont {Xiang}}, \bibinfo {author} {\bibfnamefont {Z.~P.}\ \bibnamefont {Li}}, \bibinfo {author} {\bibfnamefont {J.~M.}\ \bibnamefont {Yao}}, \bibinfo {author} {\bibfnamefont {W.~H.}\ \bibnamefont {Long}}, \bibinfo {author} {\bibfnamefont {P.}~\bibnamefont {Ring}},\ and\ \bibinfo {author} {\bibfnamefont {J.}~\bibnamefont {Meng}},\ }\href {https://doi.org/10.1103/PhysRevC.88.057301} {\bibfield  {journal} {\bibinfo  {journal} {Phys. Rev. C}\ }\textbf {\bibinfo {volume} {88}},\ \bibinfo {pages} {057301} (\bibinfo {year} {2013})}\BibitemShut {NoStop}%
\bibitem [{\citenamefont {Wang}\ \emph {et~al.}(2015)\citenamefont {Wang}, \citenamefont {Xiang}, \citenamefont {Long},\ and\ \citenamefont {Li}}]{Wang_2015}%
  \BibitemOpen
  \bibfield  {author} {\bibinfo {author} {\bibfnamefont {Z.~H.}\ \bibnamefont {Wang}}, \bibinfo {author} {\bibfnamefont {J.}~\bibnamefont {Xiang}}, \bibinfo {author} {\bibfnamefont {W.~H.}\ \bibnamefont {Long}},\ and\ \bibinfo {author} {\bibfnamefont {Z.~P.}\ \bibnamefont {Li}},\ }\href {https://doi.org/10.1088/0954-3899/42/4/045108} {\bibfield  {journal} {\bibinfo  {journal} {J. Phys. G}\ }\textbf {\bibinfo {volume} {42}},\ \bibinfo {pages} {045108} (\bibinfo {year} {2015})}\BibitemShut {NoStop}%
\bibitem [{\citenamefont {Canuto}\ \emph {et~al.}(2007)\citenamefont {Canuto}, \citenamefont {Hussaini}, \citenamefont {Quarteroni},\ and\ \citenamefont {Zang}}]{canuto2007spectral}%
  \BibitemOpen
  \bibfield  {author} {\bibinfo {author} {\bibfnamefont {C.}~\bibnamefont {Canuto}}, \bibinfo {author} {\bibfnamefont {M.~Y.}\ \bibnamefont {Hussaini}}, \bibinfo {author} {\bibfnamefont {A.}~\bibnamefont {Quarteroni}},\ and\ \bibinfo {author} {\bibfnamefont {T.~A.}\ \bibnamefont {Zang}},\ }\href {https://books.google.com/books?id=DFJB0kiq0CQC} {\emph {\bibinfo {title} {Spectral methods: fundamentals in single domains}}}\ (\bibinfo  {publisher} {Springer Berlin Heidelberg},\ \bibinfo {year} {2007})\BibitemShut {NoStop}%
\bibitem [{\citenamefont {Jiao}\ \emph {et~al.}(2021{\natexlab{a}})\citenamefont {Jiao}, \citenamefont {He}, \citenamefont {Liu}, \citenamefont {Zhang},\ and\ \citenamefont {Ho}}]{PhysRevA.104.022801}%
  \BibitemOpen
  \bibfield  {author} {\bibinfo {author} {\bibfnamefont {L.~G.}\ \bibnamefont {Jiao}}, \bibinfo {author} {\bibfnamefont {Y.~Y.}\ \bibnamefont {He}}, \bibinfo {author} {\bibfnamefont {A.}~\bibnamefont {Liu}}, \bibinfo {author} {\bibfnamefont {Y.~Z.}\ \bibnamefont {Zhang}},\ and\ \bibinfo {author} {\bibfnamefont {Y.~K.}\ \bibnamefont {Ho}},\ }\href {https://doi.org/10.1103/PhysRevA.104.022801} {\bibfield  {journal} {\bibinfo  {journal} {Phys. Rev. A}\ }\textbf {\bibinfo {volume} {104}},\ \bibinfo {pages} {022801} (\bibinfo {year} {2021}{\natexlab{a}})}\BibitemShut {NoStop}%
\bibitem [{\citenamefont {Xie}\ \emph {et~al.}(2021)\citenamefont {Xie}, \citenamefont {Jiao}, \citenamefont {Liu},\ and\ \citenamefont {Ho}}]{https://doi.org/10.1002/qua.26653}%
  \BibitemOpen
  \bibfield  {author} {\bibinfo {author} {\bibfnamefont {H.~H.}\ \bibnamefont {Xie}}, \bibinfo {author} {\bibfnamefont {L.~G.}\ \bibnamefont {Jiao}}, \bibinfo {author} {\bibfnamefont {A.}~\bibnamefont {Liu}},\ and\ \bibinfo {author} {\bibfnamefont {Y.~K.}\ \bibnamefont {Ho}},\ }\href {https://doi.org/https://doi.org/10.1002/qua.26653} {\bibfield  {journal} {\bibinfo  {journal} {Int. J. Quantum Chem.}\ }\textbf {\bibinfo {volume} {121}},\ \bibinfo {pages} {e26653} (\bibinfo {year} {2021})}\BibitemShut {NoStop}%
\bibitem [{\citenamefont {Jiao}\ \emph {et~al.}(2021{\natexlab{b}})\citenamefont {Jiao}, \citenamefont {Xie}, \citenamefont {Liu}, \citenamefont {Montgomery},\ and\ \citenamefont {Ho}}]{Jiao_2021}%
  \BibitemOpen
  \bibfield  {author} {\bibinfo {author} {\bibfnamefont {L.~G.}\ \bibnamefont {Jiao}}, \bibinfo {author} {\bibfnamefont {H.~H.}\ \bibnamefont {Xie}}, \bibinfo {author} {\bibfnamefont {A.}~\bibnamefont {Liu}}, \bibinfo {author} {\bibfnamefont {H.~E.}\ \bibnamefont {Montgomery}},\ and\ \bibinfo {author} {\bibfnamefont {Y.~K.}\ \bibnamefont {Ho}},\ }\href {https://doi.org/10.1088/1361-6455/ac259c} {\bibfield  {journal} {\bibinfo  {journal} {J. Phys. B}\ }\textbf {\bibinfo {volume} {54}},\ \bibinfo {pages} {175002} (\bibinfo {year} {2021}{\natexlab{b}})}\BibitemShut {NoStop}%
\bibitem [{\citenamefont {Press}(2007)}]{press2007numerical}%
  \BibitemOpen
  \bibfield  {author} {\bibinfo {author} {\bibfnamefont {W.~H.}\ \bibnamefont {Press}},\ }\href@noop {} {\emph {\bibinfo {title} {Numerical recipes 3rd edition: The art of scientific computing}}}\ (\bibinfo  {publisher} {Cambridge university press},\ \bibinfo {year} {2007})\BibitemShut {NoStop}%
\end{thebibliography}%

\end{document}